\documentclass[5p,twocolumn]{elsarticle}

\usepackage{amsmath}	
\usepackage{amssymb}	
\usepackage{amsthm}
\usepackage{wasysym}    
\usepackage{tikz}
\usepackage{algorithm}
\usepackage[noend]{algpseudocode}
\usepackage[colorlinks=true]{hyperref}
\usepackage[LGR,T1]{fontenc}

\newcommand{\drond}[2]{\frac{\partial #1}{\partial #2}}
\newcommand{\vect}[1]{\mathbf{\boldsymbol{#1}}}

\newcommand{\LGR}{\fontencoding{LGR}\selectfont}
\newcommand{\Latin}{\fontencoding{\encodingdefault}\selectfont}

\newcommand{\stigma}{\text{\LGR \textstigma{} \Latin}\!\!}

\newcommand{\qoppa}{\text{\LGR \textqoppa{} \Latin}\!\!}

\newcommand{\OO}{\mathcal{O}}
\newcommand{\G}{\mathcal{G}}
\newcommand{\Msun}{M_\odot}
\newcommand{\Mearth}{M_\oplus}
\newcommand{\Rearth}{R_\oplus}

\newcommand{\Rmoon}{R_{\leftmoon}}

\graphicspath{{./image/}}

\journal{New Astronomy}

\begin{document}

\begin{frontmatter}



\title{Ncorpi$\OO$N : A $\OO(N)$ software for N-body integration in collisional and fragmenting systems.}

\author[label1,label2]{J\'er\'emy COUTURIER}
\author[label2,label1]{Alice C. QUILLEN}
\author[label1,label2]{Miki NAKAJIMA}

\affiliation[label1]{organization={Dept. of Earth and Environmental Siences, University of Rochester},
	addressline={227 Hutchison Hall},
	city={Rochester},
	postcode={14627},
	state={NY},
	country={US}}
	
\affiliation[label2]{organization={Dept. of Physics and
		Astronomy, University of Rochester},
		addressline={227 Hutchison Hall},
		city={Rochester},
		postcode={14627},
		state={NY},
		country={US}}


%

\begin{abstract}

Ncorpi$\OO$N is a N-body software developed for the time-efficient integration of collisional and fragmenting systems of planetesimals or moonlets orbiting a central mass. It features a fragmentation model, based on crater scaling and ejecta models, able to realistically simulate a violent impact.

The user of Ncorpi$\OO$N can choose between four different built-in modules to compute self-gravity and detect collisions. One of these makes use of a mesh-based algorithm to treat mutual interactions in $\OO(N)$ time. Another module, much more efficient than the standard Barnes-Hut tree code, is a $\OO(N)$ tree-based algorithm called FalcON. It relies on fast multipole expansion for gravity computation and we adapted it to collision detection as well. Computation time is reduced by building the tree structure using a three-dimensional Hilbert curve. For the same precision in mutual gravity computation, Ncorpi$\OO$N is found to be up to $25$ times faster than the famous software REBOUND.

Ncorpi$\OO$N is written entirely in the C language and only needs a C compiler to run. A python add-on, that requires only basic python libraries, produces animations of the simulations from the output files. The name Ncorpi$\OO$N, reminding of a scorpion, comes from the French \textit{N-corps}, meaning N-body, and from the mathematical notation $\OO(N)$, due to the running time of the software being almost linear in the total number $N$ of moonlets. Ncorpi$\OO$N is designed for the study of accreting or fragmenting disks of planetesimal or moonlets. It detects collisions and computes mutual gravity faster than REBOUND, and unlike other N-body integrators, it can resolve a collision by fragmentation. The fast mutipole expansions are implemented up to order six to allow for a high precision in mutual gravity computation.

\end{abstract}




\end{frontmatter}


\section{Introduction}\label{sec_introduction}

Ncorpi$\OO$N is an open-source $N$-body software specialized in disk simulation, published under the GNU General Public License. It has its own website available \href{https://ncorpion.com}{here}\footnote{\label{ncorpion_superscript}\href{https://ncorpion.com}{https://ncorpion.com}} and the source code is publicly distributed on \href{https://github.com/Jeremycouturier/NcorpiON}{github}\footnote{\label{github_superscript}\href{https://github.com/Jeremycouturier/NcorpiON}{https://github.com/Jeremycouturier/NcorpiON}}. By construction, Ncorpi$\OO$N is able to integrate a system where one body (the central body) dominates in mass all the others (the orbiting bodies). The user also has the possibility of perturbing the system with a distant star, for example a star around which the central body may be orbiting if it is a planet, or a binary star if the central body is a star.
For the rest of this work, we refer to the central body of the simulation as the Earth and to the distant star as the Sun.

The development of the software began in parallel of our work on the formation of the Moon, and as such, we hereafter refer to the orbiting bodies as \textit{moonlets}. Ncorpi$\OO$N is particularly adapted to the simulation of systems where the mean free path is short, typically less than the semi-major axis, but also of systems where self-gravity plays an important role. The Moon is thought to have formed from a disk generated by a giant impact, and previous works on the formation of the Moon decide upon collision if the moonlets should bounce back or merge depending on the impact parameters (\textit{e.g.} \citet{Ida_et_al_1997}, \citet{SalmonCanup2012}), but never consider the fact that a violent collision may lead to their fragmentation. In order to address this issue, Ncorpi$\OO$N features a built-in fragmentation model that is based on numerous studies of impact and crater scaling (\citet{HolsappleHousen1986}, \citet{StewartLeinhardt2009}, \citet{HousenHolsapple2011}, \citet{LeinhardtStewart2012}, \citet{Suetsugu2018}) to properly model a violent collision. Our study of the Moon formation makes extensive use of Ncorpi$\OO$N and will be published after the present work.

Ncorpi$\OO$N comes with four different built-in modules of mutual interactions management, one of which uses the efficient fast multipole method-based falcON algorithm\footnote{An algorithm faster than the standard Barnes-Hut tree that considers cell$-$cell instead of cell$-$body interactions.} (\citet{Dehnen2002}, \citet{Dehnen2014}). Each of the four modules is able both to detect collisions and to compute self gravity. Overall, Ncorpi$\OO$N was developed with time-efficiency in mind, and its running time is almost linear in the total number $N$ of moonlets, which allows for more realistic disks to be simulated. Low-performance CPUs can be used to run Ncorpi$\OO$N.

In Sect. \ref{sec_structure}, we present the structure of the code of Ncorpi$\OO$N. In Sect. \ref{sec_mutual}, the challenging task of time-efficiently considering moonlet$-$moonlet interactions is carried out and the four built-in modules of mutual interactions management are presented.  In Sect. \ref{sec_speed}, we go over the speed performances of Ncorpi$\OO$N's four built-in modules of mutual interactions management. Finally, Sect. \ref{sec_collisions} deals with the resolution of collisions, where we present, among other things, the fragmentation model of Ncorpi$\OO$N. For convenience to the reader, we gather in Table \ref{notation} of \ref{append_notations} the notations used throughout the article. Section \ref{sec_mutual} only concerns mutual interactions between the moonlets. Other aspects of orbital dynamics, that are not moonlet$-$moonlet interactions, such as interactions with the equatorial bulge, are pushed in \ref{append_orbital} in order to prevent the paper from being too long.

Hereafter, $\OO$ denotes the center of mass of the Earth, and in a general fashion, the mass of the Earth and of the Sun are denoted by $\Mearth$ and $\Msun$, respectively. Let $N$ be the total number of moonlets orbiting the Earth and for $1\leq j\leq N$, $m_j$ is the mass of the $j^{\text{th}}$ moonlet. The geocentric inertial reference frame is $\left(\OO,\vect{i},\vect{j},\vect{k}\right)$, while the reference frame attached to the rotation of the Earth is $\left(\OO,\vect{I},\vect{J},\vect{K}\right)$, with $\vect{k}=\vect{K}$. The transformation from one to another is done through application of the rotation matrix $\vect{\Omega}$, which is the sideral rotation of the Earth. All vectors and tensors of this work are bolded, whereas their norms, as well as scalar quantities in general, are unbolded.

\section{Structure of Ncorpi$\OO$N and how to actually run a simulation}\label{sec_structure}

The \href{https://ncorpion.com}{website of Ncorpi$\OO$N}\textsuperscript{\ref{ncorpion_superscript}} features a full documentation\footnote{\href{https://ncorpion.com/\#setup}{https://ncorpion.com/\#setup}} as well as a section where the structure of the code is discussed in details\footnote{\href{https://ncorpion.com/\#structure}{https://ncorpion.com/\#structure}}. As such, it can be considered as an integral part of this work and we will refrain here from giving too much details. Instead, we stay succinct and the interested reader is invited to visit \href{https://ncorpion.com}{Ncorpi$\OO$N's website}.

Moonlets are stored in an array of structures that holds their cartesian coordinates. The moonlet structure is defined such that its size is $64$ bytes, so it fits exactly on a single cache line, which is best for cache friendliness. When arrays of dynamical size are needed, Ncorpi$\OO$N makes use of a hand-made unrolled linked list, that we call \textit{chain}. Unrolled linked lists are linked lists\footnote{A linear data structure where each node holds a value and a pointer towards the next value.} where more than one value is stored per node. Storing many values per node reduces the need for pointer dereferences and increases the locality of the storage, making unrolled linked lists significantly faster than regular linked lists.

When the mesh $\OO(N)$ algorithm is used to detect collisions and compute mutual gravity, chains are used to store the ids (in the moonlet array) of the moonlets in the different cells of the hash table. When either falcON $\OO(N)$ fast multipole method or the standard $\OO(N\ln N)$ tree code is used to detect collisions and compute mutual gravity, then chains are used to store the moonlets' ids in each cell of the octree.

The different structures used to built and manipulate the octrees are explained in the \href{https://ncorpion.com/\#structure}{website}. After the tree is built with the general construction based on pointers, it is translated into a flattree where the cells are stored in a regular array. This procedure allows for a significant CPU time to be saved (Sect. \ref{sec_Hilbert}).

Among all existing $N$-body softwares, the one closest to Ncorpi$\OO$N is REBOUND, although REBOUND does not implement falcON algorithm for mutual gravity computation and does not handle fragmentations. REBOUND is however more multi-purpose than Ncorpi$\OO$N. \href{https://teuben.github.io/nemo/man_html/gyrfalcON.1.html}{GyrfalcON} on NEMO is also similar to Ncorpi$\OO$N since it uses falcON algorithm for mutual gravity computation, but it is galaxy oriented and does not handle collisions.

The installation of Ncorpi$\OO$N from the \href{https://github.com/Jeremycouturier/NcorpiON}{github repository} is straightforward\footnote{At least on Linux and MacOS systems, we did not try to use Ncorpi$\OO$N on Windows.}. The initial conditions of the simulation, the different physical quantities, and the choice of which module is to be used for mutual interactions, is decided by the user in the parameter file of Ncorpi$\OO$N. Then, the simulation is run and an animation created from the command line. The complete documentation is provided both in the website and the github repository.

\section{Mutual interactions between the moonlets}\label{sec_mutual}

We consider in this section mutual interactions between the moonlets. The general aspects of orbital dynamics, those not related to moonlet$-$moonlet mutual interactions, are treated in \ref{append_orbital}. As long as mutual interactions between the moonlets are disregarded, the simulation runs effectively in $\OO(N)$ time. However, the moonlets can interact through collisions and mutual gravity, and managing these interactions in a naive way results in a very slow integrator. Hereafter, a mutual interaction denotes either a collision or a gravitational mutual interaction. We review in this section the four modules implemented in Ncorpi$\OO$N that can deal with mutual interactions between the moonlets, namely
\begin{itemize}
	\item $\OO(N^2)$ brute-force method.
	\item $\OO(N\ln N)$ standard tree code.
	\item $\OO(N)$ falcON fast multipole method.
	\item $\OO(N)$ mesh-grid algorithm.
\end{itemize}
Each of the four modules is able to treat both the detection of collision and the computation of self gravity (Ncorpi$\OO$N adapts Dehnen's falcON algorithm so it can also detect collisions).

\subsection{Detectiong a collision between a pair of moonlet} 

Before delving into the presentation of the four mutual interaction modules, we describe how Ncorpi$\OO$N decides if two given moonlets will collide in the upcoming timestep. Note that apart from the brute-force method, these modules rarely treat mutual interactions in a pair-wise way.

Given two moonlets with positions $\vect{r}_1$ and $\vect{r}_2$ and masses $m_1$ and $m_2$, all four modules rely on the following procedure to determine if the moonlets will be colliding during the upcoming timestep.


Let $\vect{v}_1$ and $\vect{v}_2$ be the velocities of the moonlets and $R_1$ and $R_2$ their radii. Let us denote $\Delta\vect{v}=\vect{v}_1-\vect{v}_2$ and $\Delta\vect{r}=\vect{r}_1-\vect{r}_2$. Approximating the trajectories by straight lines, we decide according to the following procedure if the moonlets will collide during the upcoming timestep. We first compute the discriminant
\begin{equation}
	\Delta = \left(\Delta\vect{r}\cdot\Delta\vect{v}\right)^2+\Delta v^2\left[\left(R_1+R_2\right)^2-\Delta r^2\right].
\end{equation}
Then, the time $\Delta t$ until the collision is given by 
\begin{equation}
	\Delta t = -\frac{\Delta\vect{r}\cdot\Delta\vect{v}+\sqrt{\Delta}}{\Delta v^2}.
\end{equation}
A collision will occur between the moonlets in the upcoming timestep if, and only if, $\Delta t\in\mathbb{R}$ and $0\leq\Delta t \leq dt$, where $dt$ is the size of the timestep. If that is the case, the collision is resolved using results from Sect. \ref{sec_collisions}.

\subsection{Brute-force $\OO(N^2)$ algorithm}

The most straightforward way of treating mutual interactions is through a brute-force algorithm where all $N(N-1)/2$ pairs of moonlets are considered. At each timestep, the mutual gravity between all pairs is computed, and the algorithm decides if a collision will occur between the two considered moonlets in the upcoming timestep. However, this greedy procedure yields a $\OO(N^2)$ time complexity, limiting the total number of moonlets to a few thousands at best on a single-core implementation (\textit{e.g.} $1000\leq N\leq2700$ in \citet{SalmonCanup2012}). 

\subsection{The mesh $\OO(N)$ algorithm}\label{simplified_sieve}

\citet{KhullerMatias1995} described in 1995 a $\OO(N)$ algorithm based on a mesh grid to find the closest pair in a set of points in the plane. Their algorithm is not completely straightforward to implement and only allows for the closest pair of moonlets to be identified. Here, we describe a mesh-based three-dimensional simplified version of their algorithm able to detect collisions in $\OO(N)$ time.
\begin{figure*}[h]
	\centering
	\includegraphics[width=\linewidth]{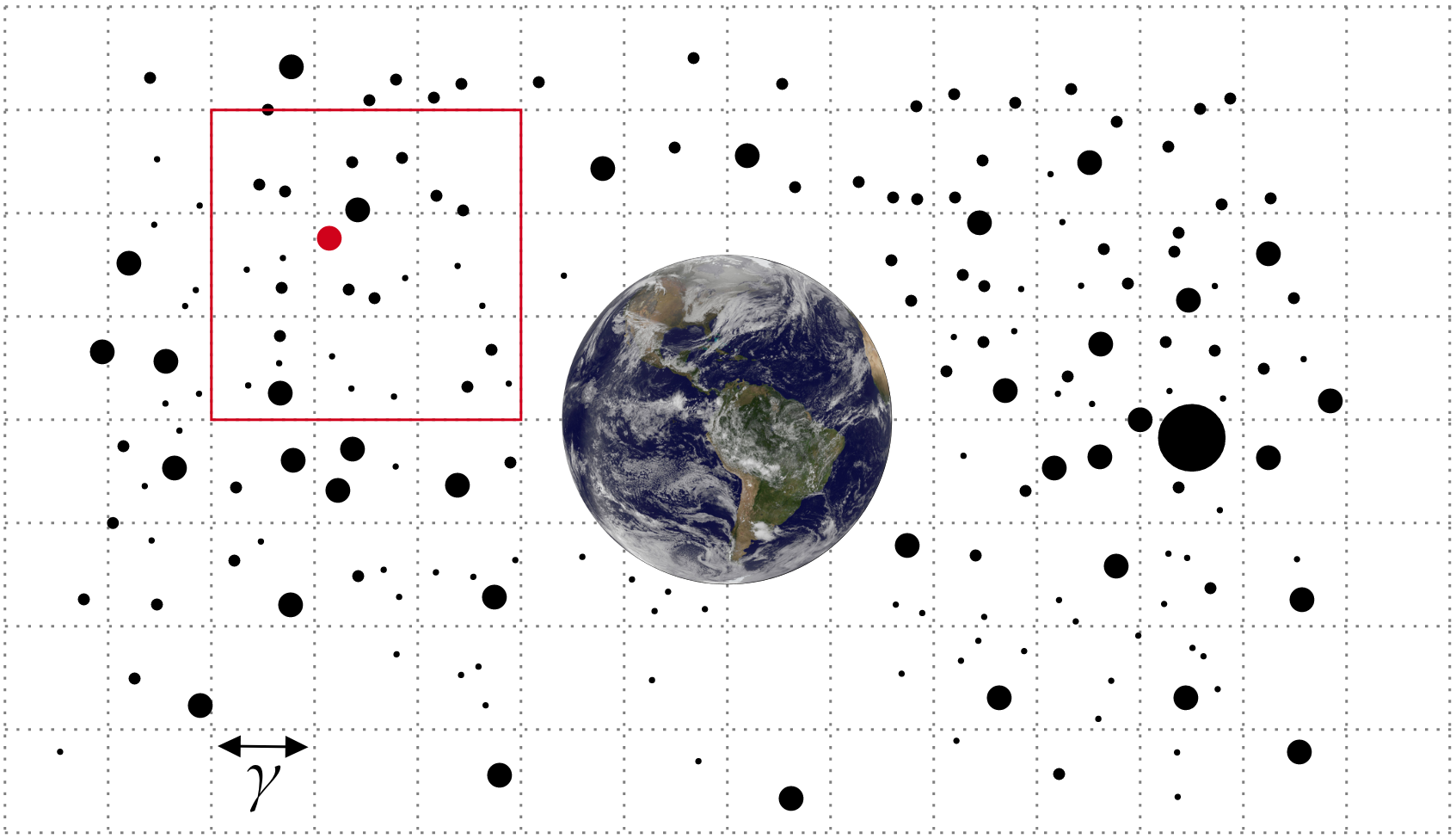}
	\caption{Schematic representation of a two-dimensional $\gamma$-mesh around the Earth. The neighbourhood of the red moonlet, defined as the cell containing it plus all the adjacent cells, is shown with a red square.}\label{fig_mesh}
\end{figure*}

For a real number $\gamma>0$, we build a $\gamma$-mesh. At each timestep, we only look for collisions between moonlets that are in each other neighborhood, and we only compute the gravitational interactions between moonlets in each other neighborhood. In Fig. \ref{fig_mesh}, we provide a schema of a $\gamma$-mesh and the definition of neighbourhood. If $\gamma$ is chosen as a function of $N$ and such that, on average, each moonlet has $\OO(1)$ moonlets in its neighborhood, then the algorithm runs in $\OO(N)$ time.

This procedure disregards gravitational interactions between moonlets not in each other neighborhood, and while the mesh algorithm is very efficient in detecting collisions, it only poorly approximates mutual gravity. In order to improve the mesh algorithm, Ncorpi$\OO$N also computes mutual gravity between any moonlets and one of the three largest moonlets. Unless the three largest moonlets account for the majority of the total moonlet mass\footnote{For example, at the end of a Moon-forming simulation.}, the mesh algorithm is poorly adapted to mutual gravity computation.

In practise in Ncorpi$\OO$N, when the mesh algorithm is used to treat mutual interactions, moonlets are put in the mesh-grid one after the other, and moonlets already populating their neighbourhood are identified. A hash table of chains is used to remember which moonlets occupy which cells of the grid. This procedure ensures that pairs are only treated once.

In order to choose the mesh-size $\gamma$, let us assume that initially, all the moonlets are located in a disk of constant aspect ratio $h/r$, at a radius $r\leq R_\text{max}$. Then they occupy a volume
\begin{equation}
	\mathcal{V}=\frac{4}{3}\pi R_\text{max}^3\sin\varsigma=\frac{4}{3}\pi R_\text{max}^3\sqrt{\frac{h^2/r^2}{1+h^2/r^2}},
\end{equation}
where $\tan\varsigma=h/r$. In order for each moonlet to have, on average, $x$ moonlets in its neighborhood, the mesh size must verify $\left(3\gamma\right)^3\leq x\mathcal{V}/N$, that is
\begin{equation}\label{gamma_crit}
	\gamma\leq\left(\frac{4\pi x}{81N}\right)^{1/3}\left(\frac{h^2/r^2}{1+h^2/r^2}\right)^{1/6}R_\text{max}.
\end{equation}
With $h/r=0.05$, $R_\text{max}=10\Rearth$, $N=10^5$ and $x=8$, this gives $\gamma=0.08526\Rearth$, or $\gamma=543.2$ km. If the $N$ moonlets have, let's say, a total mass that of the Moon, then their average radius is $R=\Rmoon/N^{1/3}$. For $N=10^5$ this gives $R=37.4$ km. The condition that the moonlets are smaller than $\gamma$ is $2R\leq\gamma$. Choosing $x=8$ and $R_\text{max}=10\Rearth$, this gives
\begin{equation}\label{varsigma_crit}
	\tan\frac{h}{r}\geq\frac{162}{\pi x}\left(\frac{R_{\leftmoon}}{R_\text{max}}\right)^3\approx0.0001307,
\end{equation}
that is, $h/r\gtrsim1.3\;10^{-4}$. Choosing for $\gamma$ the critical value given by Eq. (\ref{gamma_crit}), and for $h/r$ a value much larger than that predicted by Eq. (\ref{varsigma_crit}) ensures that most of the moonlets are smaller than the mesh-size.

In the parameter file of Ncorpi$\OO$N, the user indicates the desired number $x$ of neighbours for the simulation and Eq. (\ref{gamma_crit}) is used at the first time-step to estimate a suitable value of the mesh-size $\gamma$. Then, at each time-step, the value of $\gamma$ is updated according to the expected number $x'$ of neighbours computed at the previous timestep, in order to match the user's requirement. More precisely, if $\gamma'$ denotes the mesh-size at the previous timestep, then the new value of $\gamma$ for the current timestep is given by
\begin{equation}
	\gamma = \gamma'\left(\frac{x}{x'}\right)^{1/3}.
\end{equation}

The largest moonlets of the simulation can sometimes be larger than $\gamma$. When this happens, the corresponding moonlet is not put in the hash table but instead, mutual interactions between that moonlet and any other moonlet are treated. The user indicates in the parameter file of  Ncorpi$\OO$N the number of cells along each axis\footnote{Care must be taken to ensure that the hash table fits into the RAM.} and the minimal sidelength of the total mesh-grid, which is translated into a minimal value for the mesh-size $\gamma$.

\subsection{Tree-based algorithms}\label{sec_falcON_collision}

In this section, we present the two remaining modules of Ncorpi$\OO$N that can search for collisions or compute mutual gravity using a three-dimensional tree, or octree. The first algorithm, hereafter referred as standard tree code, was published in 1986 by \citet{BarnesHut1986} for mutual gravity computation, and adapted in 2012 by \citet{ReinLiu2012} for collision detection in REBOUND. The second algorithm, called falcON, was published in 2002 by \citet{Dehnen2002} for mutual gravity computation. We show in this section that it can be adapted to collision search as well.

Both the standard tree code and falcON use a fast multipole Taylor expansion for mutual gravity computation, and take advantage of the fact that collisions are short-range only for collision search. FalcON is significantly faster than the standard tree code at both mutual gravity computation (for the same precision) and collision detection.

\subsubsection{Tree building}
\begin{figure*}[h]
	\centering
	\includegraphics[width=\linewidth]{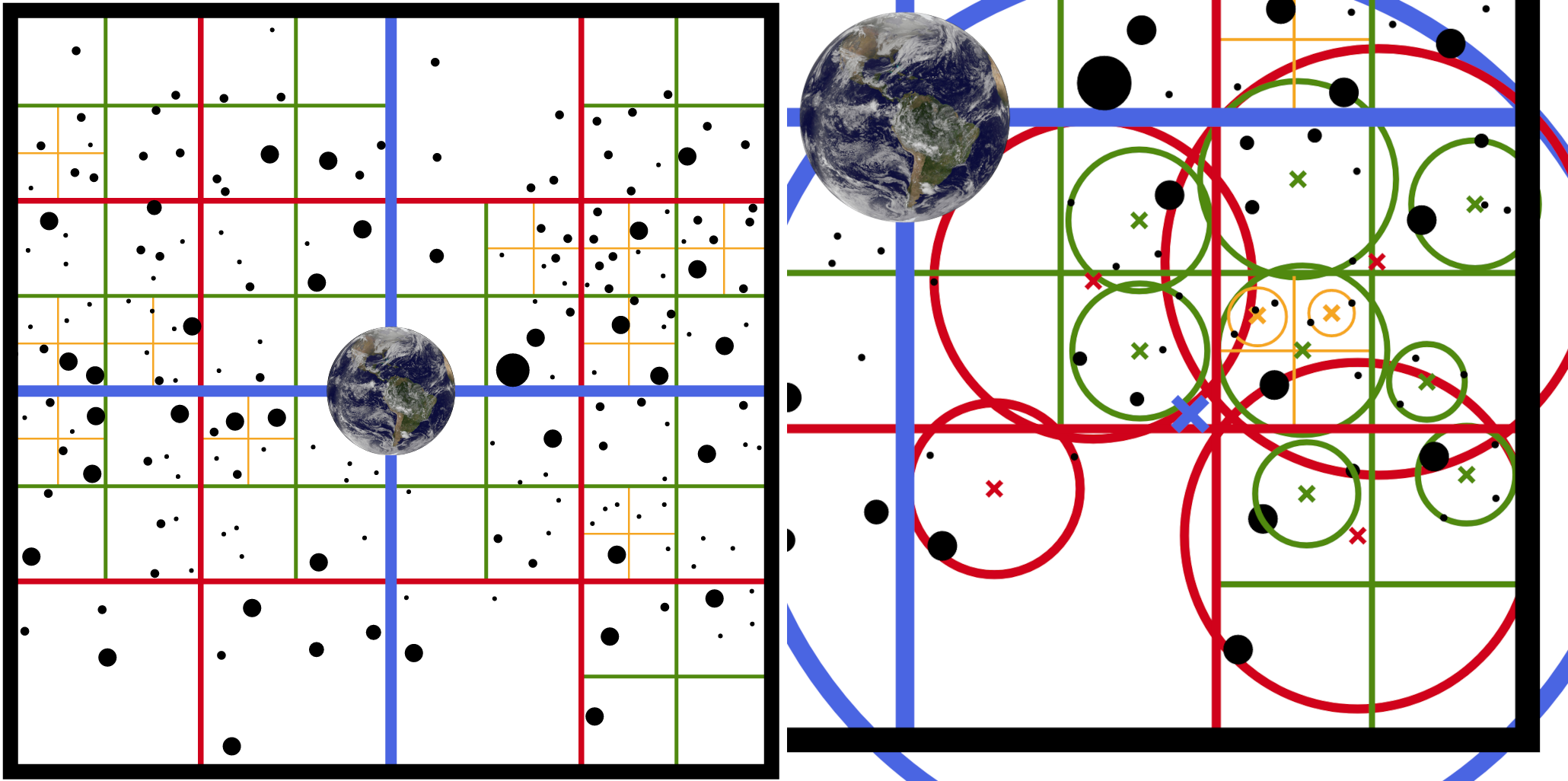}
	\caption{\textit{Left} : Schematic representation of a quadtree around the Earth built with $s=5$. The root cell is shown with a thick black square, whereas its descendants are shown with colored lines whose thicknesses decrease with their depth into the tree. The children of the root cell are blue, the grandchildren are red, etc ... \textit{Right} : A zoom-in image showing the South-East child of the root cell, its maximal radius $r_\text{max}$, and the maximal radii of its descendants. Cells with one moonlet have a zero maximal radius, and those are not shown. For clarity, the maximal circles are shown with the color and thickness of their corresponding cell, and diagonal crosses show their centers $\bar{\vect{s}}$.}\label{fig_tree}
\end{figure*}
Both algorithms use an octree, whose building procedure is not detailed here, but described in \citet{BarnesHut1986} and schematically represented in Fig. \ref{fig_tree}. Cells containing at most $s$ moonlets are not divided into children cells ($s=1$ in \citet{BarnesHut1986}). As the same tree is used for both collision search and mutual gravity computation, it is possible to build it only once per timestep to reduce the computational effort\footnote{The tree-based algorithms have a different optimal value for $s$ according to whether they are used to detect collisions or to compute mutual gravity. It could be interesting to build distinct trees with a different $s$ for collisions and gravity. Time is lost by building two trees but also saved by using the optimal $s$. We did not investigate what was best.\label{rebuilding}}.

Hereafter, we adopt the naming convention of \citet{Dehnen2002} where a node, or cell, is a cubic subdivision of the space, a child refers to a direct subnode of a node, and a descendant refers to any subcell of a cell. A leaf is a childless node and we abusively refer to the moonlets contained by a leaf as children nodes of that leaf. On the left panel of Fig. \ref{fig_tree}, we provide a schematic representation of a two-dimensional tree (quadtree) around the Earth with $s=5$.

\subsubsection{Tree climbing}

The tree climbing procedure consists in computing several quantities for all cells of the tree, recursively from the children nodes. The tree climbing procedure differs slightly for collision search or mutual gravity computation.

\subsubsection*{Collision search}

When searching for collisions, we define the center $\bar{\vect{s}}_A$ of cell $A$ as the average position of the $N_A$ children nodes it contains
\begin{equation}\label{center}
	\bar{\vect{s}}_A = \sum_{\text{child node } a\text{ of }A}\frac{\bar{\vect{s}}_a}{N_A},
\end{equation}
and its maximal and critical radii recursively as
\begin{equation}\label{rmax}
	r_{\text{max},A}=\max_{\text{child node } a\text{ of }A}\left(r_{\text{max},a} + \left|\bar{\vect{s}}_a-\bar{\vect{s}}_A\right|\right),
\end{equation}
and
\begin{equation}\label{rcrit}
	r_{\text{crit},A}=r_{\text{max},A}+\max_{\text{child node } a\text{ of }A}\;\left(r_{\text{crit},a}-r_{\text{max},a}\right).
\end{equation}
If a child node $a$ is a moonlet (that is, if $A$ is a leaf), then $\bar{\vect{s}}_a=\vect{r}_a$ is the moonlet's position, $r_{\text{max},a}=0$ and $r_{\text{crit},a}=R_a+dt\,v_a$, where $R_a$ is the moonlet's radius, $dt$ the timestep (common to all moonlets in Ncorpi$\OO$N), and $v_a$ the moonlet's scalar velocity. Starting from the leaf cells, we go up the tree and use Eqs. (\ref{center}), (\ref{rmax}) and (\ref{rcrit}) to compute recursively from the children nodes the center $\bar{\vect{s}}$, the maximal radius $r_{\text{max}}$ and the critical radius $r_{\text{crit}}$ of each cell. On the right panel of Fig. \ref{fig_tree}, we show the centers and maximal radii of the South-East child of the root cell and of its descendants.

\subsubsection*{Mutual gravity computation}

When computing mutual gravity, we define the expansion center $\vect{s}_A$ of cell $A$ of mass $M_A$ as the center of mass of the children nodes it contains
\begin{equation}\label{expansion_center}
	\vect{s}_A = \frac{1}{M_A}\sum_{\text{child node } a\text{ of }A}M_a\vect{s}_a.
\end{equation}
The maximal radius $r_{\text{max},A}$ is defined recursively in the same way as in collision search
\begin{equation}\label{rmax_2}
	r_{\text{max},A}=\max_{\text{child node } a\text{ of }A}\left(r_{\text{max},a} + \left|\vect{s}_a-\vect{s}_A\right|\right).
\end{equation}
However, the critical radius is defined differently as
\begin{equation}\label{rcrit_2}
	r_{\text{crit},A}=r_{\text{max},A}/\theta(M_A),
\end{equation}
where the opening angle $\theta(M_A)$ is given by Eq. (13) of \citet{Dehnen2002}. If a child node $a$ is a moonlet (that is, if $A$ is a leaf), then $\vect{s}_a=\vect{r}_a$ is the moonlet's position and $r_{\text{max},a}=0$. Starting from the leaf cells, we go up the tree and use Eqs. (\ref{expansion_center}), (\ref{rmax_2}) and (\ref{rcrit_2}) to compute recursively from the children nodes the expansion center $\vect{s}$, the maximal radius $r_{\text{max}}$ and the critical radius $r_{\text{crit}}$ of each cell.

For both collision search and mutual gravity computation, if the distance from the center (or expansion center) of a cell to its farthest corner is smaller than $r_{\text{max}}$, then $r_{\text{max}}$ is replaced by this distance. Two cells are said to be well-separated if the distance between their centers (or expansion centers) is larger than the sum of their critical radii, that is, if
\begin{equation}
	r_{\text{crit},A}+r_{\text{crit},B}\leq\left|\bar{\vect{s}}_A-\bar{\vect{s}}_B\right|.
\end{equation}
The same definition applies with $\vect{s}$ instead of $\bar{\vect{s}}$ for mutual gravity computation. When treating mutual gravity, the multipole moments $\vect{M}^{(n)}$ are also computed for all cells recursively from the children cells during the tree climbing. More details on doing so are provided in \ref{multipole_from_children}.

\subsubsection{Multipole expansion}\label{sec_multipole}

\begin{figure*}[h]
	\centering
	\includegraphics[width=\linewidth]{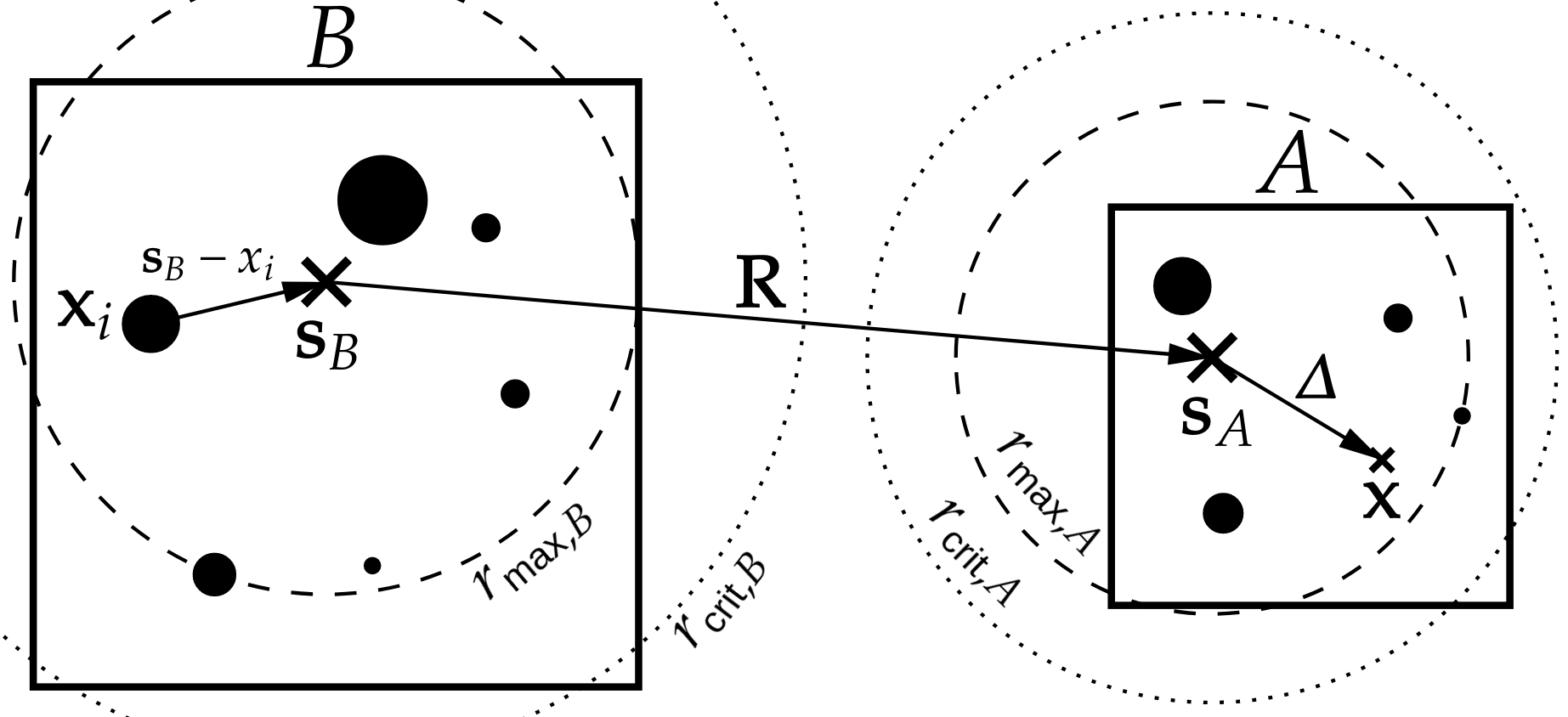}
	\caption{Multipole expansion between two interacting cells $A$ and $B$.}\label{fig_multipole}
\end{figure*}

How the octree can be used to look for collisions is trivial : Given the definition of $r_\text{crit}$, it is straightforward to verify that moonlets of cell $A$ will not collide with moonlets of cell $B$ in the upcoming timestep if $A$ and $B$ are well-separated. However, computing mutual gravitational interactions with the octree relies on multipole expansions. The existing literature provides only little insight on the mathematical framework, and we recall the theory here. In Fig. \ref{fig_multipole}, let us say that we want to compute the acceleration of moonlets of $A$ due to their gravitational interaction with moonlets of $B$. Since the critical circles do
not intersect, these two cells are well-separated. From the point of view of moonlets of $A$, moonlets of $B$ can thus be seen as a whole, and at lowest order, it is as if they were all reunited at their center of mass\footnote{This is the approximation made by \citet{BarnesHut1986} in their original description of the standard tree code, corresponding to $p=1$ in Eq. (\ref{phi_expanded}).} $\vect{s}_B$. At higher order, the mass distribution inside cell $B$ can be taken into account through the multipole moments $\vect{M}^{(n)}$ of the cell (Eq. (\ref{multipole_moment})) to reach a better precision. The gravitational potential at location $\vect{x}$ in cell $A$ due to the gravity of cell B is given by\footnote{We use here the sign convention $\ddot{\vect{x}}=\nabla\phi(\vect{x})$, in order to have $\nabla_{\vect{s}_A}\vect{C}^{(m-1)}=\vect{C}^{(m)}$ in Eq. (\ref{phi_expanded}).}
\begin{equation}\label{phi_unexpanded}
	\phi(\vect{x}) = \sum_{i\in B}\mu_i\,g(\vect{x}-\vect{x}_i),
\end{equation}
where $\mu_i=\G m_i$, $m_i$ and $\vect{x}_i$ are the mass and position of the $i^{\text{th}}$ moonlet of $B$, and $g(\vect{x})=1/x$ is the Green function of the Laplace operator. As suggested in Fig. \ref{fig_multipole}, we write
\begin{equation}
	\vect{x}-\vect{x}_i = \vect{\Delta} + \vect{R} + \vect{s}_B - \vect{x}_i,
\end{equation}
and we Taylor expand the Green function $g(\vect{x}-\vect{x}_i)$ around the cell separation $\vect{R}$ up to a certain order $p$.
\begin{equation}\label{green_expanded}
	\!\!g(\vect{x}-\vect{x}_i)=\sum_{n=0}^p\frac{(-1)^n}{n!}\vect{\nabla}^{(n)}g(\vect{R})\odot\left(\vect{x}_i-\vect{s}_B-\vect{\Delta}\right)^{(n)}\!,
\end{equation}
where a remainder of order $(r_{\text{max},A}+r_{\text{max},B})^{p+1}/R^{p+1}$ has been discarded. In this expression, the $n^{\text{th}}$ order tensor $\vect{\nabla}^{(n)}g(\vect{R})$ is the $n^{\text{th}}$ gradient of $1/R$ defined recursively as 
\begin{equation}\label{gradient}
\begin{split}
	&\!\!\!\!\nabla^{(0)}g(\vect{R})=g(\vect{R})\quad\text{and}\\
	&\!\!\!\!\!\!\left(\!\vect{\nabla}^{(n)}g(\vect{R})\!\right)^{i_1,i_2,\cdots,i_n}\!\!=\!\drond{}{R_{i_n}}\!\left(\!\vect{\nabla}^{(n-1)}g(\vect{R})\!\right)^{i_1,i_2,\cdots,i_{n-1}},\!
\end{split}
\end{equation}
with $(i_1,i_2,\cdots,i_n)\in\left\lbrace1,2,3\right\rbrace^n$ and $\vect{R}=(R_1,R_2,R_3)$. The quantity $\left(\vect{x}_i-\vect{s}_B-\vect{\Delta}\right)^{(n)}$ is the $n$-fold outer product of the vector $\vect{x}_i-\vect{s}_B-\vect{\Delta}$ with itself. The inner and outer product of two tensors are defined respectively as
\begin{equation}\label{inner_def}
	\begin{split}
	&\left(\vect{T}_1^{(n)}\odot\vect{T}_2^{(n-k)}\right)^{i_1,i_2,\cdots,i_k}=\\
	&\sum_{1\leq j_1,\cdots,j_{n-k}\leq3}T_1^{i_1,\cdots,i_k,j_1,\cdots,j_{n-k}}T_2^{j_1,\cdots,j_{n-k}},
	\end{split}
\end{equation}
and
\begin{equation}\label{outer_def}
	\left(\vect{T}_1^{(k)}\otimes\vect{T}_2^{(n-k)}\right)^{i_1,i_2,\cdots,i_n}=T_1^{i_1,\cdots,i_k}T_2^{i_{k+1},\cdots,i_n}.
\end{equation}
If we define the $n^{\text{th}}$ multipole moment of cell $B$ as the $n^{\text{th}}$ order tensor
\begin{equation}\label{multipole_moment}
	\vect{M}_B^{(n)}(\vect{s}_B)=\sum_{i\in B}\mu_i\left(\vect{x}_i-\vect{s}_B\right)^{(n)},
\end{equation}
then Eqs. (\ref{phi_unexpanded}) and (\ref{green_expanded}) yield
\begin{equation}
	\phi(\vect{x})=\sum_{n=0}^p\frac{(-1)^n}{n!}\vect{\nabla}^{(n)}g(\vect{R})\odot\vect{M}_B^{(n)}(\vect{s}_B+\vect{\Delta}).
\end{equation}
The idea is to expand \smash{$\vect{M}_B^{(n)}(\vect{s}_B+\vect{\Delta})$} as to make appear the multipole moments $\vect{M}_B^{(n)}(\vect{s}_B)$. However, since $\vect{s}_B$ and $\vect{\Delta}$ do not commute ($\vect{s}_B\otimes\vect{\Delta}\neq\vect{\Delta}\otimes\vect{s}_B$ in general), such expansion is not given by Newton's binomial. We can however use the symmetry of tensor $\vect{\nabla}^{(n)}g(\vect{R})$ to get around this difficulty. We say that a tensor $\vect{T}^{(n)}$ is symmetrical if for any permutation $\sigma$ of $\left\lbrace1,2,\cdots,n\right\rbrace$, we have
\begin{equation}\label{symmetrical_tensor}
	T^{i_1,\cdots,i_n}=T^{i_{\sigma(1)},\cdots,i_{\sigma(n)}}.
\end{equation}
Due to Schwarz rule, $\vect{\nabla}^{(n)}g(\vect{R})$ is symmetrical and we have (this is easily verified from Eqs. (\ref{inner_def}) and (\ref{outer_def}))
\begin{equation}\label{illegal_simplification}
	\begin{split}
	&\!\!\vect{\nabla}^{(n)}g(\vect{R})\odot\vect{M}_B^{(n)}(\vect{s}_B+\vect{\Delta})=\\
	&\!\!\vect{\nabla}^{(n)}g(\vect{R})\odot\left(\sum_{m=0}^n(-1)^m\binom{n}{m}\vect{\Delta}^{(m)}\otimes \vect{M}^{(n-m)}_B(\vect{s}_B)\right),\!\!\!\!\!\!
	\end{split}
\end{equation}
although Eq. (\ref{illegal_simplification}) cannot be simplified by $\vect{\nabla}^{(n)}g(\vect{R})$. Using Eq. (\ref{illegal_simplification}) and the equality
\begin{equation}
	\vect{T}_1^{(n)}\odot\left(\vect{T}_2^{(m)}\otimes\vect{T}_3^{(n-m)}\right)=\vect{T}_1^{(n)}\odot\vect{T}_2^{(m)}\odot\vect{T}_3^{(n-m)}
\end{equation}
for any symmetrical tensors $\vect{T}_1^{(n)}$, $\vect{T}_2^{(m)}$ and $\vect{T}_3^{(n-m)}$, Eq. (\ref{phi_unexpanded}) can be written (\citet{WarrenSalmon1995})
\begin{equation}\label{phi_expanded}
	\begin{split}
		&\phi(\vect{x})=\sum_{m=0}^p\frac{1}{m!}\vect{\Delta}^{(m)}\odot\vect{C}^{(m)}(\vect{s}_A),\\
		&\vect{C}^{(m)}(\vect{s}_A)=\sum_{n=0}^{p-m}\frac{(-1)^n}{n!}\vect{\nabla}^{(n+m)}g(\vect{R})\odot\vect{M}^{(n)}_B(\vect{s}_B).
	\end{split}
\end{equation}
The tensors $\vect{C}^{(0)}$, $\vect{C}^{(1)}$ and $\vect{C}^{(2)}$ are respectively the gravitational potential, the acceleration and the tidal tensor at $\vect{s}_A$ due to the gravity of cell $B$. More generally, the $\vect{C}^{(m)}$ are the interaction tensors due to the gravity of cell $B$ on the center of mass $\vect{s}_A$ of cell $A$.

In the standard tree code (Sect. \ref{sec_standard}), instead of computing interactions between cells, we compute interactions between a cell $B$ and a moonlet well-separated\footnote{The well-separation in that case is defined as $\left|\vect{x}-\vect{s}_B\right|\geq r_{\text{crit},B}$.} from $B$ at location $\vect{x}$. In that case, instead of $\vect{R}=\vect{s}_A-\vect{s}_B$, we take $\vect{R}=\vect{x}-\vect{s}_B$ and we only compute $\vect{C}^{(1)}$ in Eq. (\ref{phi_expanded}), since we are only interested in the acceleration of the moonlet.

In falcON, once the $\vect{C}^{(m)}$ due to interactions between cells of the tree have been accumulated by the tree walk (described in Sect. \ref{sec_TreeWalk}), they are passed down and accumulated by the descendants until reaching the moonlets. Since the children are not located at the expansion center $\vect{s}_0$ of their parent, their parent's $\vect{C}^{(m)}$ are translated to their own expansion center $\vect{s}_1$ (or position if the child is a moonlet). Using the equality \smash{$\nabla_{\vect{s}_0}\vect{C}^{(m-1)}(\vect{s}_0)=\vect{C}^{(m)}(\vect{s}_0)$}, this is done via a $p^{\text{th}}$ order Taylor expansion\footnote{There is a sign error in Eq. (8) of \citet{Dehnen2002}, where $\vect{s}_0-\vect{s}_1$ is written instead of $\vect{s}_1-\vect{s}_0$.}
\begin{equation}\label{shifting}
	\begin{split}
	&\vect{C}^{(m)}(\vect{s}_1)=\sum_{n=0}^{p-m}\frac{1}{n!}\nabla_{\vect{s}_0}^{(n)}\vect{C}^{(m)}(\vect{s}_0)\odot\left(\vect{s}_1-\vect{s}_0\right)^{(n)}\\
	=&\sum_{n=0}^{p-m}\frac{1}{n!}\vect{C}^{(m+n)}(\vect{s}_0)\odot\left(\vect{s}_1-\vect{s}_0\right)^{(n)}.
	\end{split}
\end{equation}
This tree descent is performed from the root cell. Once a cell has accumulated the \smash{$\vect{C}^{(m)}$} of its parent, it transmits its own \smash{$\vect{C}^{(m)}$} to its children using Eq. (\ref{shifting}), until the leaves have received the \smash{$\vect{C}^{(m)}$} from all of their ancestors. Then the accelerations \smash{$\vect{C}^{(1)}$} of the moonlets are computed from the \smash{$\vect{C}^{(m)}$} of their parent leaf using Eq. (\ref{shifting}) (\citet{Dehnen2002}, Sect. 3.2.2).

In the parameter file of Ncorpi$\OO$N, the user chooses the desired expansion order $p$ used for the multipole expansion in falcON or in the standard tree code (if the user wants to use a tree-based method for mutual interactions). In the original description of falcON by \citet{Dehnen2002}, $p$ was three, whereas $p$ was one in the original description of the standard tree code by \citet{BarnesHut1986}. Ncorpi$\OO$N allows expansion orders up to $p=6$. Since the expansion center is the center of mass, the dipole \smash{$\vect{M}^{(1)}$} vanishes by construction. Therefore, orders $p=1$ and $p=2$ are identical for the standard tree code, since only \smash{$\vect{C}^{(1)}$} is ever computed in this case.

In practise in Ncorpi$\OO$N, when using falcON, we treat the interaction of cell $B$ on cell $A$ at the same time as we treat the interaction of cell $A$ on cell $B$. The advantages of doing so are two-folds. First, we can take advantage of the relations
\begin{equation}
	\begin{split}
		&M_A\vect{C}_{B\rightarrow A}^{(p)}=(-1)^pM_B\vect{C}_{A\rightarrow B}^{(p)},\\
		&M_A\vect{C}_{B\rightarrow A}^{(p-1)}=(-1)^{p-1}M_B\vect{C}_{A\rightarrow B}^{(p-1)},\\
	\end{split}
\end{equation}
to speed up the algorithm. Second, doing so ensures that the total momentum is preserved up to machine precision, since Newton's third law is verified up to machine precision, which is not true with the standard tree code. A speed up is also achieved by noticing that the highest order multipole moment $\vect{M}^{(p)}$ only affects $\vect{C}^{(0)}$ in Eq. (\ref{phi_expanded}). Furthermore, in Eq. (\ref{shifting}), $\vect{C}^{(m)}(\vect{s}_1)$ is only affected by $\vect{C}^{(k)}(\vect{s}_0)$ for $k\geq m$. Since we are only interested in computing the accelerations of the moonlets, $\vect{C}^{(0)}$ never has to be computed for any cell, and as a consequence, the highest order multipole moment $\vect{M}^{(p)}$ is never used and does not have to be computed when climbing the tree.

Another significant speed up comes from the fact that all the manipulated tensors are symmetrical in the sense of Eq. (\ref{symmetrical_tensor}). In three-dimensional space, a symmetrical tensor of order $n$ has only $\left(n+1\right)\left(n+2\right)/2$ independent components out of the possible $3^n$. In Ncorpi$\OO$N, order $n$ tensors are therefore stored in an array of size $\left(n+1\right)\left(n+2\right)/2$, and to compute them, we only compute $\left(n+1\right)\left(n+2\right)/2$ distinct quantities. Similarly, when computing the inner product \smash{$\vect{T}_1^{(n)}\odot\vect{T}_2^{(n-k)}$}, the total number of multiplications can be reduced from $3^n$ down to only \smash{$\frac{1}{4}\left(k+1\right)\left(k+2\right)\left(n-k+1\right)\left(n-k+2\right)$} using the symmetry of the tensors.

The choice of the expansion order $p$ is an obvious parameter affecting the precision of the expansion. Another parameter is how large the critical radius $r_\text{crit}$ of a cell is, determined by Eq. (\ref{rcrit_2}). In the parameter file of Ncorpi$\OO$N, the user chooses the value of $\theta_\text{min}$, corresponding to the ratio $r_\text{max}/r_\text{crit}$ of the root cell. Then, this same ratio for the descendants of the root cell is determined by Eq. (13) of \citet{Dehnen2002}. Sensible values are $0.2\leq\theta_\text{min}\leq0.8$, and highest precisions are achieved with small values. In the standard tree code, a common practise is to consider the same $\theta$ for all cells, but here, we consider a $\theta$ dependent on the cell's mass for both falcON and the standard tree code.
\begin{figure*}[h]
	\centering
	\includegraphics[width=\linewidth]{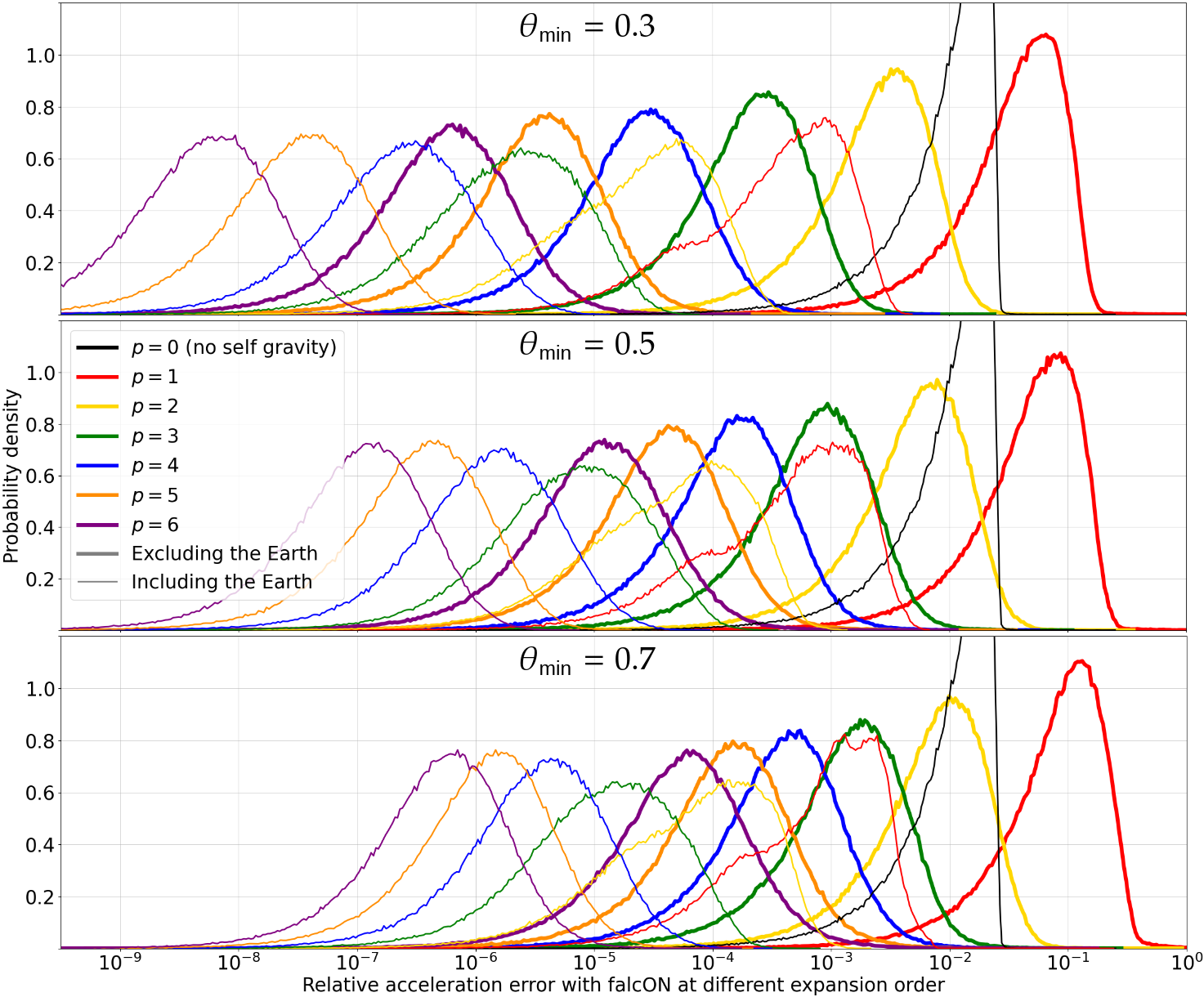}
	\caption{Distribution of the relative errors of the moonlets' accelerations with our implementation of falcON. These densities of probability were computed with $N=400\,000$ moonlets around the Earth distributed between $r=\Rearth$ and $r=10\Rearth$, each with a random mass, for a total mass $0.024\Mearth$. The relative errors are similar with the standard tree code. See text for the difference between thick and thin lines.}\label{fig_error_distribution}
\end{figure*}

In Fig. \ref{fig_error_distribution}, we plot the relative error $\left|a-\bar{a}\right|/\bar{a}$, where $a$ is the scalar acceleration computed with our implementation of falcON, whereas $\bar{a}$ is the true scalar acceleration, computed in a brute-force way. We show the probability density of the relative error, for various choices of $\theta_\text{min}$ and $p$ going from $0$ to $6$. For the thick lines, only the gravity due to the other moonlets is considered when computing $a$ and $\bar{a}$. For the thick lines, the gravity due to the Earth is also considered in the computation of $a$ and $\bar{a}$. Since the acceleration due to the Earth is computed without error\footnote{The Earth is not in the tree, and Earth$-$moonlet interactions are computed directly.}, including Earth gravity decreases greatly the relative error and the thin curves are more on the left than the thick curves in Fig. \ref{fig_error_distribution}. Choosing $p=0$ is equivalent to neglecting the mutual gravity between the moonlets, and the corresponding error distribution is plotted with a thin black line in Fig. \ref{fig_error_distribution}. 

\subsubsection{Standard tree code}\label{sec_standard}
We provide here our implementation of the standard tree code first described by \citet{BarnesHut1986}.
When an instruction differs according to whether the algorithm is used for mutual gravity computation of collision search, the instruction relative to gravity is given first in regular font, followed by the instruction relative to collision search in italic font. To treat the
interactions between all the moonlets, the procedure StandardTree is called with argument (moonlet $k$, root cell) $N$ times in a for loop going over all the moonlets once. Each call to the function is resolved in time $\OO(\ln N)$, hence the overall $\OO(N\ln N)$ time complexity. The thresholds $N_\text{cb,pre}$ and $N_\text{cb,post}$ are parameters chosen by the user. Possible values are discussed in Sect. \ref{sec_speed}.
\begin{algorithm}[h]
	\begin{algorithmic}
		\Procedure{StandardTree}{moonlet $a$, cell $B$}
		\State $N_b \gets$ number of moonlets in cell $B$ 
		\If{$N_b<N_\text{cb,pre}$}
		\State Treat the interaction between moonlet $a$ and cell $B$ brute-forcely
		\ElsIf{$a$ is well-separated from $B$}
		\State Accumulate $\vect{C}^{(1)}$ of moonlet $a$ or \textit{Do nothing}.
		\ElsIf{$N_b<N_\text{cb,post}$ or $B$ is a leaf}
		\State Treat the interaction between moonlet $a$ and cell $B$ brute-forcely
		\Else
		\ForAll{child node $b$ of $B$}
		\State\Call{StandardTree}{$a,b$}
		\EndFor
		\EndIf
		\EndProcedure
	\end{algorithmic}
\end{algorithm}

\subsubsection{FalcON : An efficient tree walk}\label{sec_TreeWalk}

We give with the algorithm TreeWalk (see below) the tree walk procedure of falcON algorithm used after the tree climbing and before the tree descent.
\begin{algorithm*}[h]
	\begin{algorithmic}
		\Procedure{TreeWalk}{cell $A$, cell $B$} \Comment{Called once on (root, root)}
		\State $(N_a,N_b) \gets$ number of moonlets in cells $A$ and $B$ 
		\If{$A=B$} \Comment{Self interaction}
		\If{$N_a \leq N_\text{cs}$ or $A$ is a leaf}
		\State Treat the interaction of cell $A$ with itself brute-forcely. \Comment{Compute gravity or \textit{search collisions} for all pairs}
		\Else
		\ForAll{pairs $(a,b)$ of children of $A$} \Comment{Up to $36$ such pairs}
		\State\Call{TreeWalk}{$a,b$}
		\EndFor
		\EndIf
		\Else \Comment{Interaction between different cells}
		\If{$N_aN_b<N_\text{cc,pre}$}
		\State Treat the interaction between cell $A$ and cell $B$ brute-forcely.
		\ElsIf{$A$ and $B$ are well-separated}
		\State For $1\leq n\leq p$, accumulate $\vect{C}^{(n)}$ for cells $A$ and $B$ (Eq. (\ref{phi_expanded})) or \textit{Do nothing}.\Comment{\textit{No collision possible}}
		\ElsIf{$N_aN_b<N_\text{cc,post}$ or both $A$ and $B$ are leaves}
		\State Treat the interaction between cell $A$ and cell $B$ brute-forcely.
		\ElsIf{$r_{\text{crit},A}>r_{\text{crit},B}$ or $B$ is a leaf} \Comment{Subdividing $A$}
		\ForAll{child node $a$ of $A$}
		\State\Call{TreeWalk}{$a,B$}
		\EndFor
		\Else \Comment{Subdividing $B$}
		\ForAll{child node $b$ of $B$}
		\State\Call{TreeWalk}{$A,b$}
		\EndFor
		\EndIf
		\EndIf
		\EndProcedure
	\end{algorithmic}
\end{algorithm*}
When a line has both regular font and italic font, only one of the two instructions is performed. The instruction in regular font is performed if falcON is used for gravity computation, whereas the instruction in italic font is applied if it is used for collision detection. Once the tree climbing is done, the tree walk procedure is called once with argument (root cell, root cell). When falcON is used for collision detection, the algorithm terminates after the tree walk. When it is used for mutual gravity computation, a tree descent stage, explained is Sect. \ref{sec_multipole}, is performed after the tree walk. The thresholds $N_\text{cs}$, $N_\text{cc,pre}$ and $N_\text{cc,post}$ are indicated by the user in the parameter file of Ncorpi$\OO$N. Possible values are discussed in Sect. \ref{sec_speed}.

In practise in Ncorpi$\OO$N, the functions TreeWalk and StandardTree are not coded recursively. Instead, we store in a stack the cell-cell interactions yet to be performed (cell-body interactions for the standard tree code) and these functions are more efficiently implemented iteratively.

\subsubsection{Peano-Hilbert order and cache efficiency}\label{sec_Hilbert}

The practical construction of a tree generally involves a structure containing relevant informations for the current cell (number of children, mass, multipole moments, etc $\cdots$), and pointers towards the children nodes, that can be either NULL or contain the address in memory of a child. Such a construction is easy to implement but yields poor memory locality (children have no reason to be next to each other in memory) and travelling in the tree requires multiple pointer dereferences. These issues are responsible for many cache misses and the processor wastes a lot of clock cycles waiting for data in memory.
\begin{figure}[h]
	\centering
	\includegraphics[width=\linewidth]{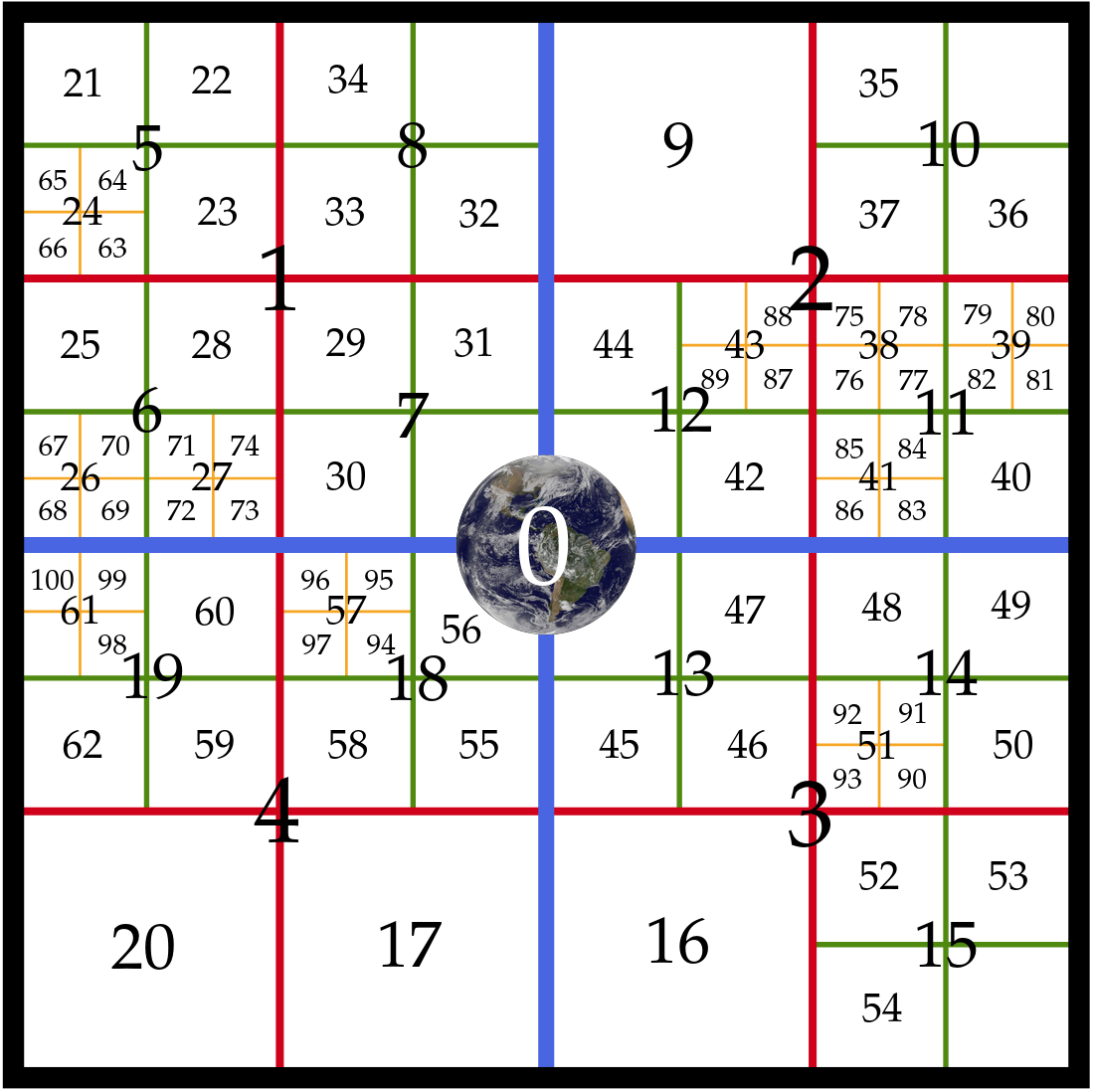}
	\caption{Hilbert order of all the cells of the tree of Fig. \ref{fig_tree}. The Hilbert order of a cell is written in its middle. Cells empty of moonlet do not exist and are not assigned an order. In practise, this tree would be stored in an array indexed from $0$ to $100$.}\label{fig_hilbert}
\end{figure}

A much better implementation can be achieved by storing the tree in a regular array, in such a way that children are contiguous in memory, and such that cells close in space are likely to be close in memory. To this aim, we use the space filling curve discovered by David Hilbert. At a given generation (or level) in the tree, cells are ordered according to a three-dimensional version of Hilbert's 1891 space filling curve, skipping non-existing cells. For illustration purposes, we show this order in two dimensions in Fig. \ref{fig_hilbert} for the tree presented in Fig. \ref{fig_tree}. For two cells $A$ and $B$, the order that we define verifies the following properties
\begin{itemize}
	\item If $\text{level}(A)>\text{level}(B)$ then $\text{order}(A)>\text{order}(B)$.
	\item If $\text{order}(A)>\text{order}(B)$ then for all child $a$ of $A$ and $b$ of $B$, $\text{order}(a)>\text{order}(b)$.
\end{itemize}
However, when building the tree, its final structure as well as the number of cells it contains are still unknown and it is not possible to build the tree in a regular array. Therefore, we use the general representation based on pointers to build the tree. Then, the final tree is copied in an array indexed by Hilbert order, hereafter called the flat tree, and the tree is freed. FalcON algorithm (climbing, walk and descent) and the standard tree code (climbing and standard tree) are performed on the flat tree.

Instead of putting the moonlets in the tree in a random order, an impressive speed-up for the tree building can be achieved by putting the moonlets in the Hilbert order of the previous timestep. We define the Hilbert order of a moonlet as the Hilbert order of its parent leaf. When the moonlets are put in the tree in the Hilbert order of the previous timestep, a spacial coherence is maintained during the tree construction, increasing the probability that the data needed by the processor are already loaded in the cache, and reducing cache misses. In Table \ref{tree_building_speedup}, we give the time taken by our CPU\textsuperscript{\ref{CPU}} to build the tree when the moonlets are added in the tree in random order and when they are added in the Hilbert order of the previous timestep. The procedure to build the tree is exactly the same in both cases, yet, cache-efficiency makes the building procedure two to three times faster.
\begin{table}[h]
	\renewcommand{\arraystretch}{1.1}
	\begin{center}
		\small
		\begin{tabular}{llllll}
			\hline
			\hline
			$N$&$2^{10}$&$2^{14}$&$2^{18}$&$2^{22}$&$2^{26}$\\
			\hline
			Random order&$75\;\,10^{-6}$&$19\;\,10^{-4}$&$0.049$&$1.5$&$45$\\
			Hilbert order&$28\;\,10^{-6}$&$7.1\;\,10^{-4}$&$0.033$&$0.65$&$13$\\
			Speed-up factor&$2.7$&$2.7$&$1.5$&$2.3$&$3.5$\\
			\hline
			\hline
		\end{tabular}
		\caption{Time (in seconds) needed to build the tree as a function of the number of moonlets $N$. When moonlets are added in random order, the tree building takes up to a factor $3.5$ longer than when they are added in the Hilbert order of the previous timestep. The subdivision threshold is $s=26$ and the moonlets are distributed between $r=2.9\Rearth$ and $r=12\Rearth$. }\label{tree_building_speedup}
	\end{center}
\end{table}

In their implementation of the standard tree code in REBOUND (\citet{ReinLiu2012}), the authors do not rebuild the tree from scratch at each timestep, but instead update it by locating moonlets that left their parent leaf. In Ncorpi$\OO$N, we prefer to build the tree from scratch at each timestep, but subsequent builds are two to three times faster than the first build thanks to Hilbert order. The authors of REBOUND do not mention the speed-up they achieved with their update procedure, and it is unknown which method is best.

\section{Numerical performances of Ncorpi$\OO$N}\label{sec_speed}

\subsection{Numerical integration}

In the parameter file of Ncorpi$\OO$N, the user chooses how moonlets interact (through collisions, mutual gravity, both of them or none of them). In case of interactions, the user also chooses how interactions should be treated (either brute-forcely, with the mesh algorithm, with falcON, or with the standard tree code). If the mesh-algorithm is used, then only mutual gravity with the neighbouring moonlets and with the three largest moonlets are taken into account. All other long range gravitational interactions between moonlets are discarded. This is generally a poor approximation, unless the three largest moonlets account for the majority of the total moonlet mass. If either falcON or the standard tree code is used, then long range mutual gravity is considered, with a precision depending on $p$ and $\theta_\text{min}$ (Fig. \ref{fig_error_distribution}).

We use a leapfrog integrator to run the numerical simulations. Depending on the method chosen for mutual interaction treatment, Ncorpi$\OO$N uses either a $SABA_1$ (\textit{half drift} $+$ \textit{kick} $+$ \textit{half drift}) or $SBAB_1$ (\textit{half kick} $+$ \textit{drift} $+$ \textit{half kick}) symplectic integrator\footnote{Whichever is faster for the given mutual interaction management method.} (\citet{LaskarRobutel2001}). When outputs do not occur at every timestep (this is generally the case for a long simulation), time is saved by combining the last step of a timestep with the first step of the next timestep, since they are identical. For example, the $SABA_1$ integrator takes in that case the form \textit{half drift} $+$ \textit{kick} $+$ \textit{drift} $+$ \textit{kick} $+$ \textit{drift} $+$ $\cdots$, until an output has to occur. When an output occurs, the last drift is undone by half (on a copy of the simulation, as to not interfere with it) and the simulation's state is written on a file. Similar considerations are valid for the $SBAB_1$ integrator. Collisions are searched and resolved during the drift phase, whereas mutual gravity is computed during the kick phase.

\subsection{Performances}

\begin{figure*}[h]
	\centering
	\includegraphics[width=\linewidth]{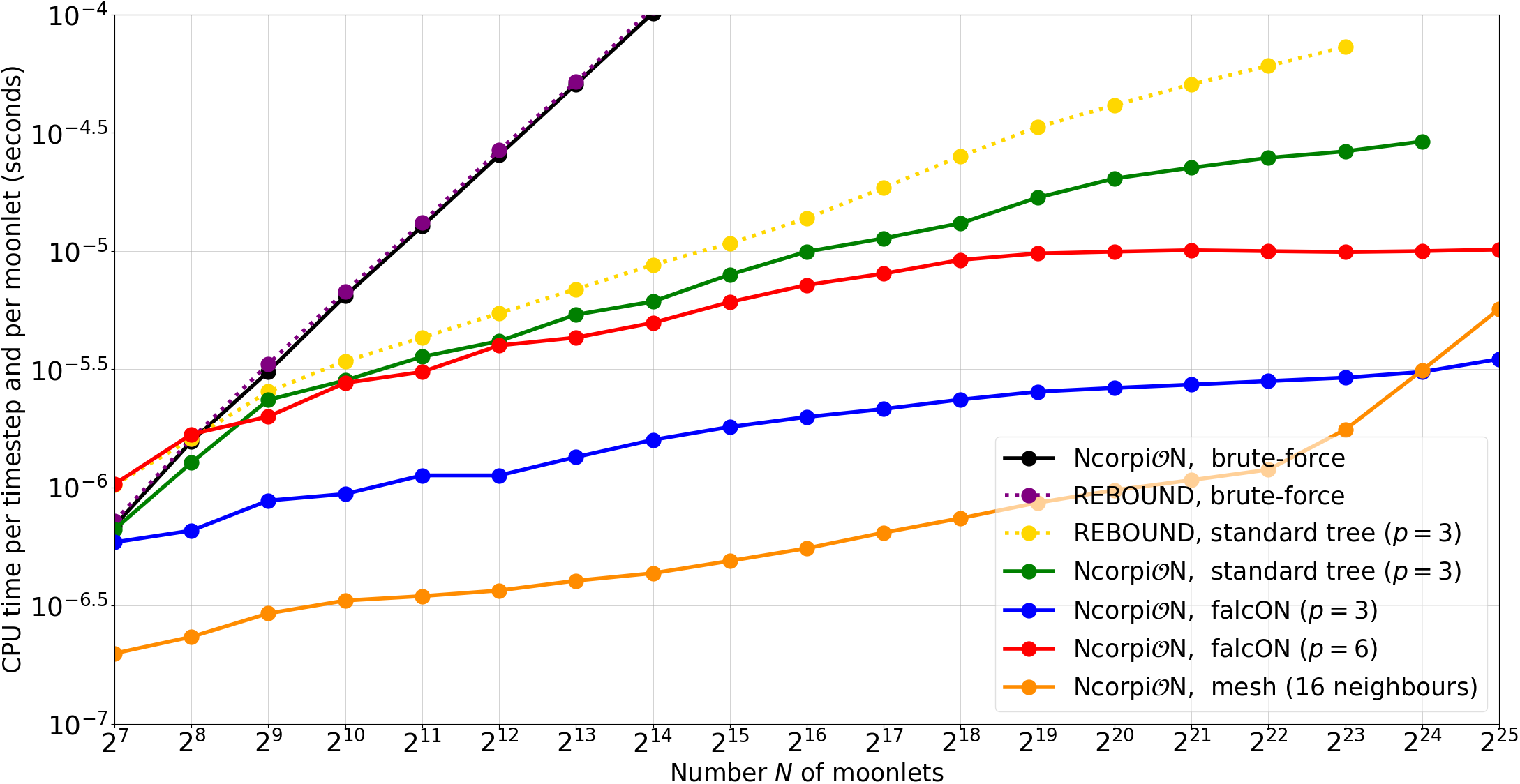}
	\caption{Time (in seconds) taken by our CPU to run one timestep (\textit{kick} $+$ \textit{drift} or \textit{drift} $+$ \textit{kick}), as a function of the number $N$ of moonlets. For clarity, the times are divided by $N$. The $N$ moonlets were given initial semi-major axes between $r=2.9\Rearth$ and $r=12\Rearth$, eccentricities between $0$ and $0.2$, and inclinations between $0^\circ$ and $10^\circ$. The mean anomaly, argument of pericenter and longitude of the ascending node were distributed uniformly between $0$ and $2\pi$. All simulations were run with the same material\textsuperscript{\ref{CPU}}, in the same conditions (all processor cores were idle except the one running).}\label{fig_speed_test}
\end{figure*}

In order to test the performances of Ncorpi$\OO$N, we ran numerical simulations with both collisions and mutual gravity, for different values of the number of moonlets $N$. In order for $N$ to be constant during a simulation, we resolved the collisions elastically. We measured the time taken by our CPU\footnote{\label{CPU}Clock : $\sim4.5$ GHz. Cache L$_1$, L$_2$, L$_3$ : $80$ KB, $1.25$ MB, $24$ MB. RAM : $32$ GB DDR5 $4800$ MT/s} to run one timestep (averaged over the first eight timesteps) with each of the four mutual interaction management methods (brute-force, falcON, standard tree code and mesh algorithm, each with the exact same initial conditions for a given $N$). We also ran the same simulations with Rein and Liu's REBOUND software (using brute-force and standard tree) in order to compare Ncorpi$\OO$N with REBOUND. In Fig. \ref{fig_speed_test}, we show the results of our tests for $2^7\leq N\leq2^{25}$.

The runs with a tree-based method (falcON or the standard tree code) were performed with $\theta_\text{min}=0.5$ ($\theta=0.5$ for REBOUND, which uses a constant $\theta$), leading to a relative error in the acceleration of the moonlets of the order of $\sim10^{-5}$ when $p=3$ and  $\sim10^{-7}$ when $p=6$ (Fig. \ref{fig_error_distribution}). The subdivision threshold $s$ is the main parameter (not precision altering) influencing the speed of the tree-based methods. For falcON, we chose $s=17$ with $p=3$ and $s=52$ with $p=6$, whereas we used $s=102$ for the standard tree code with $p=3$, as these values were optimal\footnote{Optimal only if the same tree is used for collision detection and mutual gravity computation. See footnote \ref{rebuilding}.} with our material\textsuperscript{\ref{CPU}}. For collision search, we used the thresholds $N_\text{cs}=12$, $N_\text{cc,pre}=0$, $N_\text{cc,post}=16$, $N_\text{cb,pre}=0$ and $N_\text{cb,post}=16$, whereas we used $N_\text{cs}=64$, $N_\text{cc,pre}=8$, $N_\text{cc,post}=64$, $N_\text{cb,pre}=6$ and $N_\text{cb,post}=16$ for mutual gravity computation ($p=3$). For falcON with $p=6$, we used $N_\text{cs}=N_\text{cc,pre}=256$ and $N_\text{cc,post}=512$. These thresholds influence the performances much less than the subdivision threshold $s$, and the values we report here, while among the most efficient, should not be considered as optimal.

With the brute-force method, Ncorpi$\OO$N and REBOUND turn out to run almost equally as fast (solid black and dashed purple curve in Fig. \ref{fig_speed_test}). REBOUND being slightly slower than Ncorpi$\OO$N can easily be attributed to the versatility of REBOUND, which requires larger data structures and increases the likelihood of a cache miss. On both softwares, the brute-force method is slower than any other method for $N\geq2^8$.

FalcON on Ncorpi$\OO$N turns out to be three to ten times faster than the standard tree code (blue and green curve in Fig. \ref{fig_speed_test}). Even with $p=6$ (red curve), falcON is still faster than the standard tree code with $p=3$, while also being two orders of magnitudes more precise. Our implementation of the standard tree code is also faster than that of REBOUND ($2.76$ times faster for $N=2^{23}$), which can be attributed to Ncorpi$\OO$N using a mass dependent opening angle $\theta$ (\citet{Dehnen2002}, Eq. (13)), whereas REBOUND uses a constant $\theta$. For the same precision, Ncorpi$\OO$N with falcON runs $25$ times faster than REBOUND with the standard tree code when $N=2^{23}$ (solid blue and dashed yellow curves).

Without much surprise, the mesh algorithm turns out to be the fastest method of all on Ncorpi$\OO$N for $N\leq2^{24}$, mainly due to the simplicity of its implementation. This comes at the cost of a much worse precision on the acceleration of the moonlets, since long-range gravity is ignored (unless if with one of the three largest moonlets). The dramatic increase in the running time of the method for $N\geq2^{22}$ is due to the fact that it is impossible to keep constant the average number of neighbours past a certain value of $N$. Indeed, the mesh-size $\gamma$ is attributed a minimal value to prevent the whole mesh grid (whose number of cells is constant and chosen by the user) to shrink below a certain threshold (also chosen by the user). Above this value for $N$, the average number of neighbours increases linearly instead of being constant, and the mesh algorithm behaves in $\OO\left(N^2\right)$. Therefore, falcON should be preferred to the mesh algorithm if no moonlet account for the majority of the moonlet mass, or if $N$ is too large. FalcON should always be preferred to the standard tree code. Although the mesh algorithm is faster than falcON over a full timestep due to its simplistic way of handling gravity, falcON outperforms the mesh algorithm for the drift phase, since collision search is faster with falcON. 

In Fig. \ref{fig_speed_test}, a $\OO(N)$ algorithm would have a constant curve. Therefore, none of the four mutual interaction management modules of Ncorpi$\OO$N is strictly $\OO(N)$ (although falcON is really close to it for $N\geq2^{19}$, especially at order $6$). Indeed, even if an algorithm is $\OO(N)$ in the total number of operations, when implemented on an actual CPU, the limited size of the cache is such that the proportion of cache misses increases with $N$. As a consequence, the proportion of clock cycles that the CPU spends waiting for data increases with $N$ and the time complexity ends up being slightly worse than $\OO(N)$.

\section{Resolving collisions}\label{sec_collisions}

Ncorpi$\OO$N provides several built-in ways in which collisions should be resolved. In the parameter file, the user can decide that all collisions are resolved elastically (hard-sphere collision without loss of energy), inelastically (hard-sphere collision with loss of energy), by merging the colliding moonlets together, or with the fragmentation model of Ncorpi$\OO$N, detailed in Sect. \ref{sec_frag_model}.

In this section, we consider the collision between two moonlets of masses $m_1$ and $m_2$ and radii $R_1$ and $R_2$. The positions and velocities of the moonlets, at the instant of the impact, are denoted by $\vect{r}_1$, $\vect{r}_2$, $\vect{v}_1$ and $\vect{v}_2$. We also denote
\begin{equation}\label{Deltar_Deltav}
	\Delta\vect{r}=\vect{r}_1-\vect{r}_2\quad\text{ and }\quad\Delta\vect{v}=\vect{v}_1-\vect{v}_2.
\end{equation}
In a general fashion, we refer to the largest moonlet as the target (hereafter moonlet $2$) and to the smallest one as the impactor (hereafter moonlet $1$). The impact angle is defined as\footnote{$\qoppa$ is an archaic Greek letter called qoppa.}
\begin{equation}\label{qoppa}
	\qoppa=\arcsin\left(\frac{b}{R_1+R_2}\right)=\arccos\sqrt{1-\frac{b^2}{\left(R_1+R_2\right)^2}},
\end{equation}
where $b\leq R_1+R_2$ is the impact parameter. We denote $M=m_1+m_2$ and in Ncorpi$\OO$N, all moonlets share the same density $\rho$, chosen by the user in the parameter file.

\subsection{Elastic collisions}\label{sec_elastic}

We say that a collision is elastic if it conserves both energy and momentum.
Let $\vect{v}'_1$ and $\vect{v}'_2$ be the moonlets velocities after the impact. If we write
\begin{equation}\label{velocity_prediction}
	\vect{v}'_1-\vect{v}_1=-\frac{\vect{J}}{m_1}\quad\text{ and }\quad\vect{v}'_2-\vect{v}_2=\frac{\vect{J}}{m_2},
\end{equation}
then it is immediate to verify that the total momentum is conserved, whatever the vector $\vect{J}$. Let us write
\begin{equation}
	\vect{J}=\alpha_{\text{el}}\left(\Delta\vect{r}\cdot\Delta\vect{v}\right)\Delta\vect{r},
\end{equation}
where $\alpha_{\text{el}}$ is a real number. The scalar product $\Delta\vect{r}\cdot\Delta\vect{v}$ traduces the violence of the impact, in the sense that, for a grazing collision, $\Delta\vect{r}\cdot\Delta\vect{v}=0$, while for a frontal collision, it reaches an extremum $\Delta\vect{r}\cdot\Delta\vect{v}=-\Delta r\Delta v$. The variation of kinetic energy $\Delta E$ at the impact reads
\begin{equation}
	\Delta E=\alpha_{\text{el}}\left(\Delta\vect{r}\cdot\Delta\vect{v}\right)^2\left(\frac{m_1+m_2}{2m_1m_2}\alpha_{\text{el}}\Delta r^2-1\right).
\end{equation}
At the impact, we have $\Delta r=R_1+R_2$ and the elasticity of the collision reads
\begin{equation}\label{alpha_el}
	\alpha_{\text{el}}=\frac{2m_1m_2}{\left(m_1+m_2\right)\left(R_1+R_2\right)^2}.
\end{equation}

\subsection{Inelastic collisions}\label{sec_inelastic}

The results of Sect. \ref{sec_elastic} suggest a very straightforward model for non-elastic collisions. We simply write $\vect{J}=\alpha\left(\Delta\vect{r}\cdot\Delta\vect{v}\right)\Delta\vect{r}$, and if we choose for $\alpha$ a non-zero value different from $\alpha_{\text{el}}$, then the collision in inelastic. Let us write
\begin{equation}\label{alpha}
	\alpha=\frac{fm_1m_2}{\left(m_1+m_2\right)\left(R_1+R_2\right)^2}=\frac{f}{2}\alpha_\text{el},
\end{equation}
where $f\in\mathbb{R}$. Then the variation in kinetic energy due to the impact reads
\begin{equation}\label{Delta_E}
	\Delta E=2f\left(f-2\right)\frac{m_1m_2}{m_1+m_2}\cos^2\theta\Delta v^2.
\end{equation}
To prevent an energy increase, we must consider $0\leq f\leq2$. The condition that the two moonlets gets farther away from each other after the impact reads $\Delta \vect{v}'\cdot\Delta\vect{r}\geq0$. We have 
\begin{equation}
	\Delta \vect{v}'\cdot\Delta\vect{r}=\left(1-f\right)\left(\Delta \vect{v}\cdot\Delta\vect{r}\right),
\end{equation}
and so we take $f\geq1$ to prevent the moonlets from getting closer after the collision.
Ncorpi$\OO$N's model for non-merging and non-fragmenting collisions thus relies on the parameter $f$ (indicated by the user in the parameter file of Ncorpi$\OO$N), bounded by $1\leq f\leq2$, such that values of $f$ close to $2$ correspond to almost elastic collisions, whereas values close to $1$ correspond to very inelastic collisions.

\subsection{Fragmentation and merging}\label{sec_fragmentation}

Previous studies of Moon formation (\textit{e.g.} \citet{Ida_et_al_1997}, \citet{SalmonCanup2012}) disregard the fact that, upon a violent collision, moonlets may fragment instead of just merging or bouncing back. We rely here on the existing literature about impacts and crater scaling for the velocities and sizes of the fragments in order to achieve a realistic model of fragmentation.

\subsubsection{Velocity distribution}

We first follow the impact model of \citet{HolsappleHousen1986} and \citet{HousenHolsapple2011}, based on dimensional analysis, to constrain the velocity distribution of the fragments resulting from the impact. Let $M^\star(v):=M^\star$ be the mass of fragments ejected with a velocity greater than $v$ (relative to the largest fragment). We assume the two following hypothesis:
\begin{itemize}
	\item The region of the target where material is ejected due to the impact is large enough for the impactor to be considered point-mass\footnote{This assumption is not verified for low-velocity impacts, but the moonlets merge instead of fragmenting in this case.}.
	\item The impact is violent enough to overcome both the gravity of the target and the strength of its material.
\end{itemize}
The first hypothesis clearly implies that the target is much larger than the impactor\footnote{We stress that fragmentations are poorly resolved in Ncorpi$\OO$N when the target and impactor are roughly of the same size.}, and as a consequence, the outcome of the collision does not depend on the target radius $R_2$. The second hypothesis implies that $M^\star$ does not depend on the surface gravity of the impactor, nor on the strength of its material. Another consequence of the first hypothesis is that the outcome of the impact depends on the impactor through a unique scalar quantity, called coupling parameter and defined as\footnote{\citet{Suetsugu2018} consider oblique impacts by replacing the usual $\Delta v$ by $\Delta v\cos\qoppa$, where $\qoppa$ is the impact angle.}
\begin{equation}\label{coupling}
	C=R_1\left(\Delta v\cos\qoppa\right)^\mu\rho^\nu.
\end{equation}
The exponent $\mu$ was constrained for a wide range of material assuming the accepted value\footnote{See footnote 5 of \citet{HousenHolsapple2011}.} $\nu=0.4$ and is given in Table 3 of \citet{HousenHolsapple2011}. For a non-porous target, we have $\mu=0.55$ whether it is liquid or solid, whereas $\mu=0.41$ for a rubble-pile or sand-covered target. The value of $\mu$ is to be indicated by the user of Ncorpi$\OO$N if the built-in fragmentation model is used. According to these assumptions, there exists a functional dependency of the form
\begin{equation}
	M^\star=f(C,\rho,v),
\end{equation}
that is re-written using the $\pi$-theorem as
\begin{equation}
	\frac{M^\star v^{3\mu}\rho^{3\nu-1}}{C^3}=kC_1^{3\mu},
\end{equation}
or equivalently using Eq. (\ref{coupling}) (\citet{Suetsugu2018}, Sect. 5)
\begin{equation}\label{M_star}
	\frac{M^\star(v)}{m_1}=\frac{3k}{4\pi}\left(\frac{C_1\Delta v\cos\qoppa}{v}\right)^{3\mu}.
\end{equation}
The constants $k$ and $C_1$ are given by Table 3 of \citet{HousenHolsapple2011} and are chosen in the parameter file of Ncorpi$\OO$N. For a non-porous target, we have $C_1=1.5$ (solid or liquid) and $k=0.2$ (\textit{resp}. $k=0.3$) for a liquid (\textit{resp}. solid) target. For a rubble-pile or sand-covered target, $C_1=0.55$ and $k=0.3$.

\subsubsection{Mass of the largest fragment}

Following \citet{Suetsugu2018}, we define the ejected mass $\check{m}$ as the mass unbounded to the largest fragment (or target). That is, we write
\begin{equation}\label{v_esc}
	\begin{split}
	&\check{m}=M^\star(v_\text{esc}),\text{ with }\\
	&v_\text{esc}=\sqrt{\frac{2\G M}{R}}\text{ and }R=\left(\frac{3M}{4\pi\rho}\right)^{1/3}.
	\end{split}
\end{equation}
This yields
\begin{equation}\label{m_check}
	\frac{\check{m}}{m_1}=\frac{3k}{4\pi}\left(\frac{C_1\Delta v\cos\qoppa}{v_\text{esc}}\right)^{3\mu},
\end{equation}
and the mass $\tilde{m}$ of the largest fragment is simply given by $\tilde{m}=M-\check{m}$. For a super-catastrophic collision (defined as $\tilde{m}<M/10$), Eq. (\ref{m_check}) is not valid and we use instead (\citet{LeinhardtStewart2012}, Eq. (44))
\begin{equation}\label{super_catastrophic}
	\frac{\tilde{m}}{M}=\frac{1}{10}\left(\frac{5k}{6\pi}\frac{m_1}{M}\right)^{-3/2}\left(\frac{C_1\Delta v\cos\qoppa}{v_\text{esc}}\right)^{-9\mu/2}.
\end{equation}
When a super-catastrophic collision occurs, Ncorpi$\OO$N discards the ejected mass from the simulation (assumed vaporized), and uses Eq. (\ref{super_catastrophic}) to determine the mass of the remaining moonlet.

\subsubsection{Mass of successive fragments}

Equation (\ref{m_check}) gives the mass of the largest fragment, and in this section, we give an estimate of the mass of the remaining fragments. Hereafter, the tail designates the set of all the fragments, largest excluded. \citet{LeinhardtStewart2012} fit the size distribution of the remaining fragments with
\begin{equation}\label{size_distribution}
	n(r)=Kr^{-\left(\beta+1\right)},
\end{equation}
where $n(r)dr$ is the total number of fragments with radii between $r$ and $r+dr$, and $K$ and $\beta$ are constant. Let $\tilde{m}_n$ and $r_n$ be the mass and radius of the $n^{\text{th}}$ largest fragment. We assume that all fragments are spherical with density $\rho$ and we write $\tilde{m}_1:=\tilde{m}$. The total number of fragments larger than the $n^{\text{th}}$ largest fragment is
\begin{equation}\label{n_fragments}
	n=\int_{r_n}^{+\infty}n(r)dr=\frac{K}{\beta}r_n^{-\beta},
\end{equation}
which yields \smash{$r_n=\left(n\beta/K\right)^{-1/\beta}$}. The total mass of fragments smaller than the $n^{\text{th}}$ largest fragment is given by
\begin{equation}\label{tail_mass}
	M-\sum_{k=1}^{n-1}\tilde{m}_k=\int_0^{r_n}\frac{4}{3}\pi\rho r^3n(r)dr=\frac{4\pi\rho K r_n^{3-\beta}}{3\left(3-\beta\right)}.
\end{equation}
Equations (\ref{n_fragments}) and (\ref{tail_mass}) show that a realistic description verifies $0<\beta<3$. Combining them, we obtain, for $n\geq2$, the mass of the $n^{\text{th}}$ largest fragment from the recursive expression
\begin{equation}\label{nth_fragment_recurrence}
	\tilde{m}_n=\frac{3-\beta}{n\beta}\left(M-\sum_{k=1}^{n-1}\tilde{m}_k\right).
\end{equation}
This approach predicts an infinite number of fragments, and the partial mass $\tilde{m}_1+\cdots+\tilde{m}_n$ slowly converges towards $M$ as $n$ goes to infinity. Some truncation rule on the fragment sizes has to be defined to prevent a too large number of fragments. Equation (\ref{nth_fragment_recurrence}) gives for the mass of the second largest fragment
\begin{equation}\label{tilde_m2}
	\tilde{m}_2=\frac{3-\beta}{2\beta}\check{m}:=\frac{\check{m}}{\tilde{N}}.
\end{equation}
Assuming that the tail is made up only of fragments of mass $\tilde{m}_2$, $\tilde{N}$ is defined as the number of fragments in the tail. From SPH simulations in the gravity regime, \citet{LeinhardtStewart2012} fit $\beta=2.85$, which yields $\tilde{N}=38$. In order not to overcomplicate, we assume for Ncorpi$\OO$N that all the fragments of the tail have a mass $\tilde{m}_2$. The user chooses $\tilde{N}$ and Eq. (\ref{tilde_m2}) is used to determine $\tilde{m}_2$ and the exponent $\beta$ of the power law. The fragmenting collision can be synthetized with the following schema:     
\begin{tikzpicture}[x=0.65pt,y=0.65pt,yscale=-1,xscale=1]
	\path (50,0);
	
	\draw  [line width=1.5]  (45.6,119.67) .. controls (45.6,86.16) and (72.76,59) .. (106.27,59) .. controls (139.77,59) and (166.93,86.16) .. (166.93,119.67) .. controls (166.93,153.17) and (139.77,180.33) .. (106.27,180.33) .. controls (72.76,180.33) and (45.6,153.17) .. (45.6,119.67) -- cycle ;
	\draw  [line width=1.5]  (157.6,80.2) .. controls (157.6,66.39) and (168.79,55.2) .. (182.6,55.2) .. controls (196.41,55.2) and (207.6,66.39) .. (207.6,80.2) .. controls (207.6,94.01) and (196.41,105.2) .. (182.6,105.2) .. controls (168.79,105.2) and (157.6,94.01) .. (157.6,80.2) -- cycle ;
	\draw [line width=0.75]    (106.27,119.67) -- (180.82,81.12) ;
	\draw [shift={(182.6,80.2)}, rotate = 152.66] [color={rgb, 255:red, 0; green, 0; blue, 0 }  ][line width=0.75]    (10.93,-3.29) .. controls (6.95,-1.4) and (3.31,-0.3) .. (0,0) .. controls (3.31,0.3) and (6.95,1.4) .. (10.93,3.29)   ;
	\draw [line width=0.75]    (182.6,80.2) -- (88.13,80.21) ;
	\draw [shift={(86.13,80.21)}, rotate = 359.99] [color={rgb, 255:red, 0; green, 0; blue, 0 }  ][line width=0.75]    (10.93,-3.29) .. controls (6.95,-1.4) and (3.31,-0.3) .. (0,0) .. controls (3.31,0.3) and (6.95,1.4) .. (10.93,3.29)   ;
	\draw    (134.37,80.21) .. controls (127.73,85.01) and (133.33,103.41) .. (144.43,99.93) ;
	\draw  [line width=1.5]  (230.8,119.67) .. controls (230.8,87.38) and (256.98,61.2) .. (289.27,61.2) .. controls (321.56,61.2) and (347.73,87.38) .. (347.73,119.67) .. controls (347.73,151.96) and (321.56,178.13) .. (289.27,178.13) .. controls (256.98,178.13) and (230.8,151.96) .. (230.8,119.67) -- cycle ;
	\draw  [dash pattern={on 0.84pt off 2.51pt}]  (106.27,119.67) -- (132,173.81) ;
	\draw  [dash pattern={on 0.84pt off 2.51pt}]  (182.6,80.2) -- (196,101.81) ;
	\draw  [line width=1.5]  (336.4,66.6) .. controls (336.4,56.66) and (344.46,48.6) .. (354.4,48.6) .. controls (364.34,48.6) and (372.4,56.66) .. (372.4,66.6) .. controls (372.4,76.54) and (364.34,84.6) .. (354.4,84.6) .. controls (344.46,84.6) and (336.4,76.54) .. (336.4,66.6) -- cycle ;
	\draw  [line width=1.5]  (370,91.2) .. controls (370,81.26) and (378.06,73.2) .. (388,73.2) .. controls (397.94,73.2) and (406,81.26) .. (406,91.2) .. controls (406,101.14) and (397.94,109.2) .. (388,109.2) .. controls (378.06,109.2) and (370,101.14) .. (370,91.2) -- cycle ;
	\draw  [dash pattern={on 0.84pt off 2.51pt}]  (388,91.2) -- (393.2,107.81) ;
	\draw  [line width=1.5]  (361,130.2) .. controls (361,120.26) and (369.06,112.2) .. (379,112.2) .. controls (388.94,112.2) and (397,120.26) .. (397,130.2) .. controls (397,140.14) and (388.94,148.2) .. (379,148.2) .. controls (369.06,148.2) and (361,140.14) .. (361,130.2) -- cycle ;
	\draw  [line width=1.5]  (385,161.2) .. controls (385,151.26) and (393.06,143.2) .. (403,143.2) .. controls (412.94,143.2) and (421,151.26) .. (421,161.2) .. controls (421,171.14) and (412.94,179.2) .. (403,179.2) .. controls (393.06,179.2) and (385,171.14) .. (385,161.2) -- cycle ;
	\draw  [line width=1.5]  (345,173.2) .. controls (345,163.26) and (353.06,155.2) .. (363,155.2) .. controls (372.94,155.2) and (381,163.26) .. (381,173.2) .. controls (381,183.14) and (372.94,191.2) .. (363,191.2) .. controls (353.06,191.2) and (345,183.14) .. (345,173.2) -- cycle ;
	\draw  [dash pattern={on 0.84pt off 2.51pt}]  (363,173.2) -- (368.2,189.81) ;
	\draw  [dash pattern={on 0.84pt off 2.51pt}]  (403,161.2) -- (408.2,177.81) ;
	\draw  [dash pattern={on 0.84pt off 2.51pt}]  (379,130.2) -- (384.2,146.81) ;
	\draw  [dash pattern={on 0.84pt off 2.51pt}]  (354.4,66.6) -- (359.6,83.21) ;
	\draw  [dash pattern={on 0.84pt off 2.51pt}] (333.2,74.6) .. controls (306.2,51.6) and (356.2,2.6) .. (400.2,45.6) .. controls (444.2,88.6) and (443.2,175.6) .. (426.2,198.6) .. controls (409.2,221.6) and (295.2,227.6) .. (328.2,182.6) .. controls (361.2,137.6) and (360.2,97.6) .. (333.2,74.6) -- cycle ;
	\draw  [dash pattern={on 0.84pt off 2.51pt}]  (289.27,119.67) -- (306,175.6) ;
	\draw [line width=0.75]  [dash pattern={on 4.5pt off 4.5pt}]  (220.5,16) -- (220.5,180.6)(217.5,16) -- (217.5,180.6) ;
	\draw   (112,192.8) .. controls (112,189.38) and (114.78,186.6) .. (118.2,186.6) -- (282.8,186.6) .. controls (286.22,186.6) and (289,189.38) .. (289,192.8) -- (289,211.4) .. controls (289,214.82) and (286.22,217.6) .. (282.8,217.6) -- (118.2,217.6) .. controls (114.78,217.6) and (112,214.82) .. (112,211.4) -- cycle ;
	\draw    (65,24) -- (160,24) ;
	\draw    (289,24) -- (373,24) ;
	
	\draw (132.4,104.2) node [anchor=north west][inner sep=0.75pt]    {$\mathbf{\Delta r}$};
	\draw (108.27,65.0) node [anchor=north west][inner sep=0.75pt]    {$\mathbf{\Delta v}$};
	\draw (62.8,49) node [anchor=north west][inner sep=0.75pt]    {$m_{2}$};
	\draw (170.6,41.4) node [anchor=north west][inner sep=0.75pt]    {$m_{1}$};
	\draw (120,87.4) node [anchor=north west][inner sep=0.75pt]    {$\qoppa$};
	\draw (120,137.4) node [anchor=north west][inner sep=0.75pt]  [font=\small]  {$R_{2}$};
	\draw (187,76) node [anchor=north west][inner sep=0.75pt]  [font=\small]  {$R_{1}$};
	\draw (353.93,59.93) node [anchor=north west][inner sep=0.75pt]  [font=\scriptsize]  {$\tilde{R}_{2}$};
	\draw (383.8,58.6) node [anchor=north west][inner sep=0.75pt]  [font=\small]  {$\tilde{m}_{2}$};
	\draw (362.93,166.73) node [anchor=north west][inner sep=0.75pt]  [font=\scriptsize]  {$\tilde{R}_{2}$};
	\draw (353.6,34.2) node [anchor=north west][inner sep=0.75pt]  [font=\small]  {$\tilde{m}_{2}$};
	\draw (392.93,12.49) node [anchor=north west][inner sep=0.75pt]  [font=\small,rotate=-45.32]  {$\check{m} =\tilde{N} \ \tilde{m}_{2}$};
	\draw (332.6,184.2) node [anchor=north west][inner sep=0.75pt]  [font=\small]  {$\tilde{m}_{2}$};
	\draw (397.6,181.2) node [anchor=north west][inner sep=0.75pt]  [font=\small]  {$\tilde{m}_{2}$};
	\draw (396.6,113.2) node [anchor=north west][inner sep=0.75pt]  [font=\small]  {$\tilde{m}_{2}$};
	\draw (378.90,122.73) node [anchor=north west][inner sep=0.75pt]  [font=\scriptsize]  {$\tilde{R}_{2}$};
	\draw (387.6,83.73) node [anchor=north west][inner sep=0.75pt]  [font=\scriptsize]  {$\tilde{R}_{2}$};
	\draw (402.63,153.4) node [anchor=north west][inner sep=0.75pt]  [font=\scriptsize]  {$\tilde{R}_{2}$};
	\draw (254,49.4) node [anchor=north west][inner sep=0.75pt]    {$\tilde{m}$};
	\draw (64,7) node [anchor=north west][inner sep=0.75pt]   [align=left] {Before impact};
	\draw (287,7) node [anchor=north west][inner sep=0.75pt]   [align=left] {After impact};
	\draw (118.2,194) node [anchor=north west][inner sep=0.75pt]    {$m_{1}+m_{2}=\tilde{m}+\check{m}=M$};
	\draw (298,130.4) node [anchor=north west][inner sep=0.75pt]    {$\tilde{R}$};
	\draw (314.85,43.89) node [anchor=north west][inner sep=0.75pt]  [font=\footnotesize,rotate=-302.82] [align=left] {tail};
	
\end{tikzpicture}

\subsubsection{The fragmentation model of Ncorpi$\OO\!$N}\label{sec_frag_model}

The built-in fragmentation model of Ncorpi$\OO$N proceeds as follow. In the parameter file, the user defines a mass threshold \smash{$m^{(0)}$} (We use $m^{(0)}=5\times10^{-9}\Mearth$ for our work about the Moon formation), such that:
\begin{itemize}
	\item If \smash{$\tilde{m}_2\geq m^{(0)}$}, then the two moonlets are broken into $\tilde{N}+1$ pieces, where the largest fragment has a mass $\tilde{m}=M-\check{m}$ given by Eq. (\ref{m_check}), and the $\tilde{N}$ other fragments have a mass $\tilde{m}_2$ given by Eq. (\ref{tilde_m2}). 
	\item If \smash{$\tilde{m}_2< m^{(0)}\leq\check{m}$}, then the tail is made up of one unique fragment of mass \smash{$\tilde{N}\tilde{m}_2=\check{m}$}.
	\item If \smash{$\check{m}<m^{(0)}$}, then the collision results in a merger.
	\item If $\tilde{m}<M/10$, then the impact is super-catastrophic. Equation (\ref{super_catastrophic}) is used and the tail is discarded.
	\item The largest fragment is given velocity $\tilde{\vect{v}}$ and position $\tilde{\vect{r}}$ determined in Sect. \ref{sec_momentum}, whereas the $\tilde{N}$ (\textit{resp}. one) other fragments have velocities $\tilde{\vect{v}}_1,\;\tilde{\vect{v}}_2,\;\cdots,\;\tilde{\vect{v}}_{\tilde{N}}$ (\textit{resp}. $\check{\vect{v}}$) determined in Sect. \ref{sec_velocities}.
\end{itemize}

\subsubsection{Velocities of fragments}\label{sec_velocities}

We now estimate the velocities of the fragments after the impact, using Eq. (\ref{M_star}) for $M^\star(v)$. We define
\begin{equation}\label{m_star}
	m^\star(v) := -\frac{dM^\star}{dv} = 3\mu\check{m}\frac{1}{v}\left(\frac{v_{\text{esc}}}{v}\right)^{3\mu},
\end{equation}
where $m^\star(v)dv$ is the mass of fragments with speeds relative to the largest fragment in the range $\left[v,v+dv\right]$. Since all $\tilde{N}$ fragments of the tail are unbounded to the largest fragment, the slowest of these is made up of particles having been ejected with velocities between $v_\text{esc}$ and some velocity $u_1$. More generally, the $k^{\text{th}}$ fastest fragment of the tail has a velocity $\tilde{v}_k'$ with respect to the largest fragment given by
\begin{equation}\label{m2vk}
	\tilde{m}_2\tilde{v}_k'=\int_{u_{k-1}}^{u_k}m(v)vdv,
\end{equation}
where $u_0=v_\text{esc}$, $u_{\tilde{N}}=+\infty$ and for all $k\leq\tilde{N}$, $u_{k-1}<\tilde{v}_k'<u_k$. The speeds $u_k$ are found by writing
\begin{equation}\label{z_k_determination}
	\!\!\!\!\tilde{m}_2=\int_{u_{k-1}}^{u_k}m(v)dv=\check{m}\left[\left(\frac{v_\text{esc}}{u_{k-1}}\right)^{3\mu}-\left(\frac{v_\text{esc}}{u_k}\right)^{3\mu}\right].
\end{equation}
If we define $z_k=\left(v_\text{esc}/u_k\right)^{3\mu}$, then Eq. (\ref{z_k_determination}) yields $z_{k-1}-z_k=\tilde{m}_2/\check{m}$, that is
\begin{equation}\label{z_k}
	z_k=1-\frac{k}{\tilde{N}},\quad\text{for }0\leq k\leq\tilde{N}.
\end{equation}
Injecting Eq. (\ref{z_k}) into Eq. (\ref{m2vk}), we obtain the scalar velocity of the $k^{\text{th}}$ fastest fragment of the tail
\begin{equation}\label{tilde_vk}
	\frac{\tilde{v}_k'}{v_\text{esc}}=\frac{\tilde{N}}{\stigma}\left[\left(1-\frac{k-1}{\tilde{N}}\right)^{\stigma}-\left(1-\frac{k}{\tilde{N}}\right)^{\stigma}\right],
\end{equation}
where\footnote{$\stigma$ is an archaic Greek letter called stigma.} $\stigma=\left(3\mu-1\right)/3\mu$. Surprisingly enough, these speeds are independent of $\Delta v$, suggesting that a high impact velocity means more fragmentation but does not translate into a faster ejecta. When the tail is made up of one unique fragment, its scalar velocity is given by
\begin{equation}\label{check_v}
	\check{v}'=\frac{1}{\check{m}}\int_{v_\text{esc}}^{+\infty}m(v)vdv=\frac{v_\text{esc}}{\stigma}.
\end{equation}
The existing literature gives little insight on the directions of fragments following an impact (\citet{Suo_et_al_2023} give some constraints but their work is limited to impacts on granular media in an intermediate regime between gravity and strength), and our model here is arbitrary. The speeds of the tail's fragments are given a direction with respect to the largest fragment in the following way. We give to the $k^{\text{th}}$ fragment of the tail the position
\begin{equation}\label{r_k_tilde}
	\tilde{\vect{r}}_k'=\Delta\vect{r}+2p_k\tilde{R}_2\vect{u}+2q_k\tilde{R}_2\vect{v},
\end{equation}
and the speed
\begin{equation}\label{v_k_tilde}
	\tilde{\vect{v}}_k'=\tilde{v}_k'\frac{\tilde{\vect{r}}_k'}{\tilde{r}_k'}=\tilde{v}_k'\frac{\tilde{\vect{r}}_k'}{\sqrt{\left(R_1+R_2\right)^2+4\tilde{R}_2^2\left(p_k^2+q_k^2\right)}},
\end{equation}
where $\left(p_k,q_k\right)\in\mathbb{Z}^2$, $\tilde{v}_k$ is given by Eq. (\ref{tilde_vk}) and the vectors $\vect{u}$ and $\vect{v}$ are defined by
\begin{equation}
	\vect{v}=\frac{\vect{\Delta r}\times\vect{\Delta v}}{\Delta r\Delta v\sin\qoppa}\quad\text{and}\quad\vect{u}=\frac{\vect{v}\times\vect{\Delta r}}{R_1+R_2}.
\end{equation}
If the collision is nearly frontal, then the vector $\vect{v}$ is ill-defined. In that case we take for $\vect{v}$ any unit vector orthogonal to $\Delta\vect{r}$. With $\tilde{N}=15$ (or $\beta=45/17$), $-1\leq p_k\leq3$ and\footnote{This choice ensures that more fragments are ejected forward than backward, which sounds intuitive.} $-1\leq q_k\leq1$, the fragmented moonlets would look like the following schema

\begin{tikzpicture}[x=1pt,y=1pt,yscale=-1,xscale=1]
	\path (20,30);
	
	\draw  [line width=1.5]  (66.67,143.49) .. controls (66.67,111.55) and (92.55,85.67) .. (124.49,85.67) .. controls (156.42,85.67) and (182.31,111.55) .. (182.31,143.49) .. controls (182.31,175.42) and (156.42,201.31) .. (124.49,201.31) .. controls (92.55,201.31) and (66.67,175.42) .. (66.67,143.49) -- cycle ;
	\draw  [color={rgb, 255:red, 128; green, 128; blue, 128 }  ,draw opacity=1 ][fill={rgb, 255:red, 74; green, 144; blue, 226 }  ,fill opacity=1 ] (192.39,100.79) .. controls (192.39,97.21) and (195.3,94.3) .. (198.88,94.3) .. controls (202.47,94.3) and (205.37,97.21) .. (205.37,100.79) .. controls (205.37,104.37) and (202.47,107.28) .. (198.88,107.28) .. controls (195.3,107.28) and (192.39,104.37) .. (192.39,100.79) -- cycle ;
	\draw  [color={rgb, 255:red, 128; green, 128; blue, 128 }  ,draw opacity=1 ][fill={rgb, 255:red, 74; green, 144; blue, 226 }  ,fill opacity=1 ] (184.66,90.09) .. controls (184.66,86.51) and (187.56,83.6) .. (191.14,83.6) .. controls (194.73,83.6) and (197.63,86.51) .. (197.63,90.09) .. controls (197.63,93.67) and (194.73,96.58) .. (191.14,96.58) .. controls (187.56,96.58) and (184.66,93.67) .. (184.66,90.09) -- cycle ;
	\draw  [color={rgb, 255:red, 128; green, 128; blue, 128 }  ,draw opacity=1 ][fill={rgb, 255:red, 74; green, 144; blue, 226 }  ,fill opacity=1 ] (200.66,110.67) .. controls (200.66,107.09) and (203.56,104.18) .. (207.14,104.18) .. controls (210.73,104.18) and (213.63,107.09) .. (213.63,110.67) .. controls (213.63,114.26) and (210.73,117.16) .. (207.14,117.16) .. controls (203.56,117.16) and (200.66,114.26) .. (200.66,110.67) -- cycle ;
	\draw  [color={rgb, 255:red, 128; green, 128; blue, 128 }  ,draw opacity=1 ][fill={rgb, 255:red, 74; green, 144; blue, 226 }  ,fill opacity=1 ] (177.07,79.26) .. controls (177.07,75.67) and (179.98,72.77) .. (183.56,72.77) .. controls (187.14,72.77) and (190.05,75.67) .. (190.05,79.26) .. controls (190.05,82.84) and (187.14,85.74) .. (183.56,85.74) .. controls (179.98,85.74) and (177.07,82.84) .. (177.07,79.26) -- cycle ;
	\draw  [color={rgb, 255:red, 128; green, 128; blue, 128 }  ,draw opacity=1 ][fill={rgb, 255:red, 74; green, 144; blue, 226 }  ,fill opacity=1 ] (169.74,68.59) .. controls (169.74,65.01) and (172.64,62.1) .. (176.23,62.1) .. controls (179.81,62.1) and (182.72,65.01) .. (182.72,68.59) .. controls (182.72,72.17) and (179.81,75.08) .. (176.23,75.08) .. controls (172.64,75.08) and (169.74,72.17) .. (169.74,68.59) -- cycle ;
	\draw  [color={rgb, 255:red, 128; green, 128; blue, 128 }  ,draw opacity=1 ][fill={rgb, 255:red, 74; green, 144; blue, 226 }  ,fill opacity=1 ] (164.74,71.51) .. controls (164.74,67.92) and (167.64,65.02) .. (171.23,65.02) .. controls (174.81,65.02) and (177.72,67.92) .. (177.72,71.51) .. controls (177.72,75.09) and (174.81,77.99) .. (171.23,77.99) .. controls (167.64,77.99) and (164.74,75.09) .. (164.74,71.51) -- cycle ;
	\draw  [color={rgb, 255:red, 128; green, 128; blue, 128 }  ,draw opacity=1 ][fill={rgb, 255:red, 74; green, 144; blue, 226 }  ,fill opacity=1 ] (158.57,75.51) .. controls (158.57,71.92) and (161.48,69.02) .. (165.06,69.02) .. controls (168.64,69.02) and (171.55,71.92) .. (171.55,75.51) .. controls (171.55,79.09) and (168.64,81.99) .. (165.06,81.99) .. controls (161.48,81.99) and (158.57,79.09) .. (158.57,75.51) -- cycle ;
	\draw  [color={rgb, 255:red, 128; green, 128; blue, 128 }  ,draw opacity=1 ][fill={rgb, 255:red, 74; green, 144; blue, 226 }  ,fill opacity=1 ] (171.99,82.09) .. controls (171.99,78.51) and (174.89,75.6) .. (178.48,75.6) .. controls (182.06,75.6) and (184.97,78.51) .. (184.97,82.09) .. controls (184.97,85.67) and (182.06,88.58) .. (178.48,88.58) .. controls (174.89,88.58) and (171.99,85.67) .. (171.99,82.09) -- cycle ;
	\draw  [color={rgb, 255:red, 128; green, 128; blue, 128 }  ,draw opacity=1 ][fill={rgb, 255:red, 74; green, 144; blue, 226 }  ,fill opacity=1 ] (195.57,114.01) .. controls (195.57,110.42) and (198.48,107.52) .. (202.06,107.52) .. controls (205.64,107.52) and (208.55,110.42) .. (208.55,114.01) .. controls (208.55,117.59) and (205.64,120.49) .. (202.06,120.49) .. controls (198.48,120.49) and (195.57,117.59) .. (195.57,114.01) -- cycle ;
	\draw  [color={rgb, 255:red, 128; green, 128; blue, 128 }  ,draw opacity=1 ][fill={rgb, 255:red, 74; green, 144; blue, 226 }  ,fill opacity=1 ] (187.16,103.78) .. controls (187.16,100.19) and (190.06,97.29) .. (193.64,97.29) .. controls (197.23,97.29) and (200.13,100.19) .. (200.13,103.78) .. controls (200.13,107.36) and (197.23,110.27) .. (193.64,110.27) .. controls (190.06,110.27) and (187.16,107.36) .. (187.16,103.78) -- cycle ;
	\draw  [color={rgb, 255:red, 128; green, 128; blue, 128 }  ,draw opacity=1 ][fill={rgb, 255:red, 74; green, 144; blue, 226 }  ,fill opacity=1 ] (189.74,117.76) .. controls (189.74,114.17) and (192.64,111.27) .. (196.23,111.27) .. controls (199.81,111.27) and (202.72,114.17) .. (202.72,117.76) .. controls (202.72,121.34) and (199.81,124.24) .. (196.23,124.24) .. controls (192.64,124.24) and (189.74,121.34) .. (189.74,117.76) -- cycle ;
	\draw  [color={rgb, 255:red, 128; green, 128; blue, 128 }  ,draw opacity=1 ][fill={rgb, 255:red, 74; green, 144; blue, 226 }  ,fill opacity=1 ] (182.49,106.76) .. controls (182.49,103.17) and (185.39,100.27) .. (188.98,100.27) .. controls (192.56,100.27) and (195.47,103.17) .. (195.47,106.76) .. controls (195.47,110.34) and (192.56,113.24) .. (188.98,113.24) .. controls (185.39,113.24) and (182.49,110.34) .. (182.49,106.76) -- cycle ;
	\draw  [color={rgb, 255:red, 128; green, 128; blue, 128 }  ,draw opacity=1 ][fill={rgb, 255:red, 74; green, 144; blue, 226 }  ,fill opacity=1 ] (165.66,86.51) .. controls (165.66,82.92) and (168.56,80.02) .. (172.14,80.02) .. controls (175.73,80.02) and (178.63,82.92) .. (178.63,86.51) .. controls (178.63,90.09) and (175.73,92.99) .. (172.14,92.99) .. controls (168.56,92.99) and (165.66,90.09) .. (165.66,86.51) -- cycle ;
	\draw  [color={rgb, 255:red, 128; green, 128; blue, 128 }  ,draw opacity=1 ][fill={rgb, 255:red, 74; green, 144; blue, 226 }  ,fill opacity=1 ] (179.41,93.42) .. controls (179.41,89.84) and (182.31,86.93) .. (185.89,86.93) .. controls (189.48,86.93) and (192.38,89.84) .. (192.38,93.42) .. controls (192.38,97.01) and (189.48,99.91) .. (185.89,99.91) .. controls (182.31,99.91) and (179.41,97.01) .. (179.41,93.42) -- cycle ;
	\draw  [color={rgb, 255:red, 128; green, 128; blue, 128 }  ,draw opacity=1 ][fill={rgb, 255:red, 74; green, 144; blue, 226 }  ,fill opacity=1 ] (173.82,96.51) .. controls (173.82,92.92) and (176.73,90.02) .. (180.31,90.02) .. controls (183.89,90.02) and (186.8,92.92) .. (186.8,96.51) .. controls (186.8,100.09) and (183.89,102.99) .. (180.31,102.99) .. controls (176.73,102.99) and (173.82,100.09) .. (173.82,96.51) -- cycle ;
	\draw [line width=0.75]    (124.49,143.49) -- (191.91,104.77) ;
	\draw [shift={(193.64,103.78)}, rotate = 150.13] [color={rgb, 255:red, 0; green, 0; blue, 0 }  ][line width=0.75]    (10.93,-4.9) .. controls (6.95,-2.3) and (3.31,-0.67) .. (0,0) .. controls (3.31,0.67) and (6.95,2.3) .. (10.93,4.9)   ;
	\draw   (167.58,118.11) .. controls (167.58,117.5) and (168.08,117) .. (168.69,117) .. controls (169.3,117) and (169.8,117.5) .. (169.8,118.11) .. controls (169.8,118.72) and (169.3,119.22) .. (168.69,119.22) .. controls (168.08,119.22) and (167.58,118.72) .. (167.58,118.11) -- cycle ;
	\draw [color={rgb, 255:red, 128; green, 128; blue, 128 }  ,draw opacity=1 ]   (171.23,71.51) -- (188.69,45.26) ;
	\draw [shift={(189.8,43.6)}, rotate = 123.65] [color={rgb, 255:red, 128; green, 128; blue, 128 }  ,draw opacity=1 ][line width=0.75]    (10.93,-4.9) .. controls (6.95,-2.3) and (3.31,-0.67) .. (0,0) .. controls (3.31,0.67) and (6.95,2.3) .. (10.93,4.9)   ;
	\draw [color={rgb, 255:red, 128; green, 128; blue, 128 }  ,draw opacity=1 ]   (178.48,82.09) -- (201.8,56.09) ;
	\draw [shift={(203.13,54.6)}, rotate = 131.89] [color={rgb, 255:red, 128; green, 128; blue, 128 }  ,draw opacity=1 ][line width=0.75]    (10.93,-4.9) .. controls (6.95,-2.3) and (3.31,-0.67) .. (0,0) .. controls (3.31,0.67) and (6.95,2.3) .. (10.93,4.9)   ;
	\draw [color={rgb, 255:red, 128; green, 128; blue, 128 }  ,draw opacity=1 ]   (185.89,93.42) -- (213.26,70.87) ;
	\draw [shift={(214.8,69.6)}, rotate = 140.51] [color={rgb, 255:red, 128; green, 128; blue, 128 }  ,draw opacity=1 ][line width=0.75]    (10.93,-4.9) .. controls (6.95,-2.3) and (3.31,-0.67) .. (0,0) .. controls (3.31,0.67) and (6.95,2.3) .. (10.93,4.9)   ;
	\draw [color={rgb, 255:red, 128; green, 128; blue, 128 }  ,draw opacity=1 ]   (193.64,103.78) -- (224.41,85.62) ;
	\draw [shift={(226.13,84.6)}, rotate = 149.45] [color={rgb, 255:red, 128; green, 128; blue, 128 }  ,draw opacity=1 ][line width=0.75]    (10.93,-4.9) .. controls (6.95,-2.3) and (3.31,-0.67) .. (0,0) .. controls (3.31,0.67) and (6.95,2.3) .. (10.93,4.9)   ;
	\draw [color={rgb, 255:red, 128; green, 128; blue, 128 }  ,draw opacity=1 ]   (202.06,114.01) -- (232.94,101.99) ;
	\draw [shift={(234.8,101.27)}, rotate = 158.74] [color={rgb, 255:red, 128; green, 128; blue, 128 }  ,draw opacity=1 ][line width=0.75]    (10.93,-4.9) .. controls (6.95,-2.3) and (3.31,-0.67) .. (0,0) .. controls (3.31,0.67) and (6.95,2.3) .. (10.93,4.9)   ;
	
	\draw (146.67,130.07) node [anchor=north west][inner sep=0.75pt]    {$\Delta\vect{r}$};
	\draw (171.02,48.27) node [anchor=north west][inner sep=0.75pt]  [font=\tiny,color={rgb, 255:red, 128; green, 128; blue, 128 }  ,opacity=1 ]  {$\tilde{\vect{v}}_{k}$};
	\draw (183.68,57.6) node [anchor=north west][inner sep=0.75pt]  [font=\tiny,color={rgb, 255:red, 128; green, 128; blue, 128 }  ,opacity=1 ]  {$\tilde{\vect{v}}_{k}$};
	\draw (194.02,68.6) node [anchor=north west][inner sep=0.75pt]  [font=\tiny,color={rgb, 255:red, 128; green, 128; blue, 128 }  ,opacity=1 ]  {$\tilde{\vect{v}}_{k}$};
	\draw (204.02,81.27) node [anchor=north west][inner sep=0.75pt]  [font=\tiny,color={rgb, 255:red, 128; green, 128; blue, 128 }  ,opacity=1 ]  {$\tilde{\vect{v}}_{k}$};
	\draw (212.68,95.27) node [anchor=north west][inner sep=0.75pt]  [font=\tiny,color={rgb, 255:red, 128; green, 128; blue, 128 }  ,opacity=1 ]  {$\tilde{\vect{v}}_{k}$};
	\draw (80.67,79.4) node [anchor=north west][inner sep=0.75pt]    {$\tilde{m}$};
\end{tikzpicture}
\vspace{3mm}

While all fragments of the tail are unbounded to the largest fragment, there is no reason why the fragments of the tail should be unbounded to one another. Depending on their relative velocities, some pairs will be unbounded while some others will not. With $\tilde{N}=15$ the scalar velocities range from $\tilde{v}_1\approx1.02\,v_\text{esc}$ to $\tilde{v}_{15}\approx13.1\,v_\text{esc}$, and considering that $v_\text{esc}$ is the escape velocity at the surface of a complete merger, it is clear that most pairs, if not all, will be unbounded.

\subsubsection{Conservation of momentum and angular momentum}\label{sec_momentum}

Choosing the position $\tilde{\vect{r}}$ and velocity $\tilde{\vect{v}}$ of the largest fragment completes the definition of Ncorpi$\OO$N's fragmentation model. Indeed, the positions and speeds of the tail's moonlets are then given in the geocentric reference frame $\left(\OO,\vect{i},\vect{j},\vect{k}\right)$ by
\begin{equation}\label{rk_vk}
	\tilde{\vect{r}}_k=\tilde{\vect{r}}_k'+\tilde{\vect{r}},\quad\tilde{\vect{v}}_k=\tilde{\vect{v}}_k'+\tilde{\vect{v}}.
\end{equation}
When the tail is reunited into a single moonlet, its position and speed are $\check{\vect{r}}=\check{\vect{r}}'+\tilde{\vect{r}}$ and $\check{\vect{v}}=\check{\vect{v}}'+\tilde{\vect{v}}$, where $\check{\vect{r}}'$ and $\check{\vect{v}}'$ are defined by Eqs. (\ref{r_k_tilde}) and (\ref{v_k_tilde}) with $p_1=q_1=0$. Let
\begin{equation}\label{vcm_rcm}
	\vect{v}_\text{cm}=\frac{m_1}{M}\vect{v_1}+\frac{m_2}{M}\vect{v_2}\quad\text{and}\quad\vect{r}_\text{cm}=\frac{m_1}{M}\vect{r_1}+\frac{m_2}{M}\vect{r_2}
\end{equation}
be the velocity and position of the center of mass of the colliding pair. We define
\begin{equation}\label{angular_momentum}
	\vect{G}=m_1\vect{r}_1\times\vect{v}_1+m_2\vect{r}_2\times\vect{v}_2
\end{equation}
the angular momentum of the pair at the collision. For a merger, the conservation of the angular momentum (\textit{resp}. the momentum) reads $M\tilde{\vect{r}}\times\tilde{\vect{v}}=\vect{G}$ (\textit{resp}. $\tilde{\vect{v}}=\vect{v}_\text{cm}$). It is interesting to notice that it is impossible to preserve both the momentum and the angular momentum at the collision without considering the spin. Indeed, the conservation of the angular momentum implies that $\tilde{\vect{v}}$ is orthogonal to $\vect{G}$. However, from
\begin{equation}
	M\tilde{\vect{v}}\cdot\vect{G}=M\vect{v}_\text{cm}\cdot\vect{G}=m_1m_2\vect{v}_2\cdot\Delta\vect{r}\times\Delta\vect{v},
\end{equation}
we conclude that it is possible to conserve both the momentum and the angular momentum only if $\Delta\vect{r}\times\Delta\vect{v}=\vect{0}$, or equivalently, only if the collision is frontal ($\qoppa=0$). For oblique collisions, the only way to conserve both is to take into account the spin of the moonlets. However, taking into account the spin complexifies the treatment of collisions as well as the numerical implementation and slows down the code. Therefore, NcorpiON does not implement the spin and if the user chooses to use the fragmentation model or to resolve all collisions by merging, then it must be decided if the momentum or the angular momentum should be preserved upon impact. If falcON is used to treat mutual interactions, then it makes more sense to preserve the momentum upon collision, since by construction, falcON preserves the total momentum when computing mutual gravity, but does not preserve the total angular momentum.

When the colliding moonlets merge, the momentum is conserved simply by taking $\tilde{\vect{v}}=\vect{v}_\text{cm}$, whereas we achieve the conservation of the momentum with $\tilde{\vect{v}}=\vect{v}_\text{cm}-\check{m}\check{\vect{v}}'/M$ when the tail is reunited into one unique moonlet. Finally, when a full fragmentation occurs, we conserve the total momentum with
\begin{equation}
	\tilde{\vect{v}}=\vect{v}_\text{cm}-\sum_{j=1}^{\tilde{N}}\frac{\tilde{m}_2}{M}\tilde{\vect{v}}_k'.
\end{equation}
Conserving the angular momentum is not as straightforward and we present our model for doing so in \ref{angular_momentum_conservation}. Requiring the conservation of the momentum is not enough to constrain $\tilde{\vect{r}}$, and we use the value of $\tilde{\vect{r}}$ determined in \ref{angular_momentum_conservation} even when the user decides to conserve the momentum, in order to be as close as possible from the conservation of the total angular momentum.






\section{Conclusions}\label{sec_conclusion}

We have presented with this paper a novel $N$-body software, faster than existing $N$-body integrators on a single core implementation. Unlike other similar softwares, Ncorpi$\OO$N is able to treat a fragmentation subsequent to a violent collision. Mutual interactions (collisions and self-gravity) can be treated with four different modules, whose time complexities range from $\OO(N)$ to $\OO(N^2)$. Using falcON module for mutual interactions, Ncorpi$\OO$N is found to be $25$ times faster than the software REBOUND when $N=2^{23}$, for the same precision in mutual gravity computation.

Ncorpi$\OO$N is very adapted to simulations of satellites or planet formation, and we are currently using it to better understand the formation of the Moon from a protolunar disk around the Earth, following the giant impact between the proto-Earth and Theia. The results of this study will constitute another paper, that will be published afterwards.

Ncorpi$\OO$N has its own \href{https://ncorpion.com}{website}\textsuperscript{\ref{ncorpion_superscript}} and is distributed freely on the following \href{https://github.com/Jeremycouturier/NcorpiON}{github repository}\textsuperscript{\ref{github_superscript}}. Both these resources provide extensive documentation and the website also provides with a detailed overview of the structure of Ncorpi$\OO$N's code.

This software was written with time efficiency in mind and aims to be as CPU-efficient and cache-friendly as possible. As such, we believe it is among the fastest single-core $N$-body codes for large $N$, if not the fastest\footnote{\href{https://teuben.github.io/nemo/man_html/gyrfalcON.1.html}{GyrfalcON} on NEMO could be faster, since it also uses falcON. However, it does not handle collisions or fragmentations.}. However, unlike other softwares, Ncorpi$\OO$N lacks a parallelized version. Even though REBOUND is found to be significantly slower than Ncorpi$\OO$N on a single-core run, it would outperform Ncorpi$\OO$N if heavily parallelized. Therefore, we plan to upgrade Ncorpi$\OO$N to a parallelized version in the future, as well as to make it a multi-purpose $N$-body integrator (by removing the need for a massive central mass).

Ncorpi$\OO$N has its own fragmentation module that relies on crater scaling and ejecta models to come up with a realistic outcome for violent collisions between moonlets. However, this model makes assumptions (\textit{e.g.} impactor much smaller than target) that can be hard to reconcile with the reality of a simulation. Furthermore, the direction of the fragments is chosen arbitrarily after a fragmentation, and these issues could reduce the actual degree of realism of Ncorpi$\OO$N's fragmentation model. 

Beyond planetary or satellite formation, disks of debris are also observed by stellar occultation around some trans-Neptunian object like the dwarf planet Haumea (\citet{Ortiz_et_al_2017}), or a smaller-sized body called Quaoar (\citet{Morgado_et_al_2023}). Both these objects feature rings located outside of their Roche radius, and Ncorpi$\OO$N could be a relevant tool to understand what mechanisms prevent the rings' material from accreting.

\section*{Acknowledgements}
J\'er\'emy Couturier thanks Walter Dehnen for helpful transatlantic discussions about the intricacies of FalcON algorithm as well as Hanno Rein for help with REBOUND. This work was partly supported by NASA grants 80NSSC19K0514 and 80NSSC21K1184. Partial funding was also provided by the Center for Matter at Atomic Pressures (CMAP), the National Science Foundation (NSF) Physics Frontier Center under Award PHY-2020249, and EAR-2237730 by NSF.  Any opinions, findings, conclusions or recommendations expressed in this material are those of the authors and do not necessarily reflect those of the National Science Foundation. This work was also supported in part by the Alfred P. Sloan Foundation under grant number G202114194.

\section*{Author contributions and competing interests statement}
J.C. had the idea of the Ncorpi$\OO$N software and wrote its code, whereas A.Q. and M.N. provided ideas and insight. The authors declare that they do not have competing interests.


\appendix

\section{Notations}\label{append_notations}

\begin{table*}[h]
	\begin{center}
		\begin{tabular}{lr|lr|lr|lr}
			\hline
			\hline
			Notation&Definition&Notation&Definition&Notation&Definition&Notation&Definition\\
			\hline
			bold&vector or tensor&$\dot{}$&$d/dt$&$N$&$\!\!\!\!\!\!\!\!\!\!\!\!\!\!\!\!\!\!\!\!\!\!\!\!\!\!\!$number of moonlets&$\Rearth$&Earth radius\\
			
			$\Mearth$&Earth mass&$\Msun$&Sun mass&$\Rmoon$&Moon radius&$m_j$&moonlet mass\\
			
			$\vect{r}$&moonlet position&$\vect{v}$&moonlet speed&$\mathcal{G}$&$\!\!\!\!\!\!\!\!\!\!\!\!\!\!\!\!\!\!\!\!\!\!\!$gravitational constant&$\zeta$&geoid altitude\\
			
			$Y_{lm}$&$\!\!\!\!\!\!\!\!\!\!\!$spherical harmonic&$\vect{\Omega}$&Earth rotation&$\Omega_\text{c}$&\smash{$\left(\G\Mearth/\Rearth^3\right)^{1/2}$}&$J_2$&$\!\!\!\!\!\!\!\!\!\!\!\!\!2^{\text{nd}}\,$zonal$\,$harmonic\\
			
			$P_2(z)$&$\frac{1}{2}\left(3z^2-1\right)$&$\gamma$&mesh-size&$x$&$\!\!\!\!\!\!\!\!\!\!\!\!\!\!\!\!\!\!$number$\,$of$\,$neighbours&$s$&$\!\!\!\!\!\!\!\!\!\!\!\!\!\!\!\!\!\!$subdivision$\,$threshold\\
			
			$\bar{\vect{s}}_A$&Eq. (\ref{center})&$r_\text{max}$&$\!\!\!\!\!\!\!\!\!\!\!\!\!\!\!\!\!\!$Eqs. (\ref{rmax}) \& (\ref{rmax_2})&$r_\text{crit}$&$\!\!\!\!\!\!\!\!\!\!\!\!\!\!\!\!\!\!$Eqs. (\ref{rcrit}) \& (\ref{rcrit_2})&$\vect{s}_A$&Eq. (\ref{expansion_center})\\
			
			$\theta$&$\!\!\!\!\!\!\!\!\!\!\!\!\!\!\!\!\!\!$opening angle&\smash{$\vect{M}^{(n)}$}&Eq. (\ref{multipole_moment})&$\mu_i$&$\G m_i$&$\vect{\Delta},\vect{R}$&Fig. \ref{fig_multipole}\\
			
			$g(\vect{z})$&$1/z$&\smash{$\vect{\nabla}^{(n)}\!g(\vect{R})\!\!\!\!\!\!\!\!\!\!\!\!\!\!\!\!\!\!\!\!\!\!\!$}&Eq. (\ref{gradient})&$p$&$\!\!\!\!\!\!\!\!\!\!\!\!\!\!\!\!\!\!$expansion order&$\odot$&Eq. (\ref{inner_def})\\
			
			$\otimes$&Eq. (\ref{outer_def})&\smash{$\vect{C}^{(n)}$}&Eq. (\ref{phi_expanded})&\smash{$\vect{T}^{(n)}$}&$\!\!\!\!\!\!\!\!\!\!\!\!\!\!\!\!\!\!$arbitrary tensor&$N_\text{cc,cb,cs}$&$\!\!\!\!\!\!\!$Sect.$\,$\ref{sec_standard}$\,$\&$\,$\ref{sec_TreeWalk}\\
			
			$\Delta\vect{r},\Delta\vect{v}$&Eq. (\ref{Deltar_Deltav})&$\qoppa$&Eq. (\ref{qoppa})&$b$&$\!\!\!\!\!\!\!\!\!\!\!\!\!\!\!\!\!\!$impact parameter&$\rho$&$\!\!\!\!\!\!\!\!\!\!\!\!\!\!\!\!\!\!$moonlet density\\
			
			$\alpha$&$\!\!\!\!\!\!\!\!\!\!\!\!\!\!\!\!\!\!$Eqs. (\ref{alpha_el}) \& (\ref{alpha})&$f$&Eq. (\ref{alpha})&$C,\mu,\nu\!\!\!\!\!\!$&Eq. (\ref{coupling})&$M^\star(v)$&Eq. (\ref{M_star})\\
			
			$k,C_1$&$\!\!\!\!\!\!\!\!\!\!\!\!\!\!\!\!\!\!$below Eq. (\ref{M_star})&$v_\text{esc}$&Eq. (\ref{v_esc})&$\check{m}$&Eq. (\ref{m_check})&$\tilde{m}$&$\!\!\!\!\!\!\!\!\!\!\!\!\!\!\!\!\!\!$below Eq. (\ref{m_check})\\
			
			$\beta$&Eq. (\ref{size_distribution})&$\tilde{m}_j$&Eq. (\ref{nth_fragment_recurrence})&\smash{$\tilde{N}$}&Eq. (\ref{tilde_m2})&\smash{$m^{(0)}$}&Sect. \ref{sec_frag_model}\\
			
			\smash{$\tilde{\vect{r}},\tilde{\vect{v}}$}&$\!\!\!\!\!\!\!\!\!\!\!\!\!\!\!\!\!\!$\ref{angular_momentum_conservation}&$m^\star(v)$&Eq. (\ref{m_star})&$\stigma$&$\left(3\mu-1\right)/3\mu$&\smash{$\tilde{\vect{r}}_k',\tilde{\vect{v}}_k'$}&$\!\!\!\!\!\!\!\!\!\!\!\!\!\!\!\!\!\!$Eqs (\ref{r_k_tilde})$\,$\&$\,$(\ref{v_k_tilde})\\
			
			\smash{$\tilde{\vect{r}}_k,\tilde{\vect{v}}_k$}&Eq. (\ref{rk_vk})&$\vect{r}_\text{cm},\vect{v}_\text{cm}$&Eq. (\ref{vcm_rcm})&$\vect{G}$&Eq. (\ref{angular_momentum})&$\vect{\Lambda}$&$\vect{s}_B-\vect{s}_b$\\
			
			$M$&$m_1+m_2$&\smash{$\tilde{\vect{g}},\tilde{\vect{s}},\tilde{\vect{u}}\!\!\!\!\!\!\!\!\!\!\!\!$}&Eq. (\ref{gsu})&&&\\
			
			\hline
			\hline
		\end{tabular}
		\caption{Notations used in this paper, ordered roughly by first appearance (top to bottom). Notations used right after their definition only are not included here. $\stigma$ (stigma) and $\qoppa$ (qoppa) are archaic Greek letters.}\label{notation}
	\end{center}
\end{table*}
We gather for convenience all the notations used throughout this work in Table \ref{notation}.

\section{General orbital dynamics}\label{append_orbital}

This appendix focuses on aspects of orbital dynamics that are not moonlet$-$moonlet interactions (treated in Sect. \ref{sec_mutual}).

\subsection{Interactions with the center of mass of the Earth}

We consider here the gravitational interactions between the moonlets and the center of mass of the Earth. Let $\vect{r}$ be the position of a moonlet in the geocentric reference frame. Its gravitational potential per unit mass reads
\begin{equation}
	V=-\frac{\G\Mearth}{r},
\end{equation}
where $\G$ is the gravitational constant. The moonlet's acceleration is given by
\begin{equation}\label{acceleration}
	\ddot{\vect{r}}=-\nabla_\vect{r}V,
\end{equation}
that is,
\begin{equation}
	\ddot{\vect{r}}=-\frac{\G\Mearth}{r^3}\vect{r}.
\end{equation}
Ncorpi$\OO$N uses dimensionless units such that $\Rearth=\Mearth=1$ and $\G=4\pi^2$. The choice $\G=4\pi^2$, instead of the more common $\G=1$, ensures that the unit of time is the orbital period at Earth surface. The user can change this in the parameter file.

\subsection{Earth flattening and interactions with the equatorial bulge}

The Earth is not exactly a sphere, and under its own rotation, it tends to take an ellipsoidal shape. The subsequent redistribution of mass modifies its gravitational field, affecting the moonlets. Let $\zeta(\theta,\varphi)$ be the altitude of the geoid of the Earth, where
\begin{equation}
	\begin{split}
		&X=r\sin\theta\cos\varphi,\\
		&Y=r\sin\theta\sin\varphi,\\
		&Z=r\cos\theta,
	\end{split}
\end{equation}
is the relation between the cartesian and spherical coordinates of $\left(\OO,\vect{I},\vect{J},\vect{K}\right)$. If $\Rearth$ denotes the mean radius of the Earth, then the geoid is generally defined as the only equipotential surface such that
\begin{equation}\label{geoid_definition}
	\int_0^{2\pi}\int_0^{\pi}\zeta(\theta,\varphi)\sin\theta d\theta d\varphi=\Rearth,
\end{equation}
that is, as the only equipotential surface whose average height is the mean radius. Expanding the geoid over the spherical harmonics as
\begin{equation}
	\begin{split}
		&\zeta(\theta,\varphi)=\Rearth\left[1+h(\theta,\varphi)\right],\text{ and}\\
		&h(\theta,\varphi)=\sum_{l=2}^{+\infty}\sum_{m=-l}^l\epsilon_{lm}Y_{lm}(\theta,\varphi)\\
	\end{split}
\end{equation}
satisfies Eq. (\ref{geoid_definition}). For reference, the definition of the spherical harmonics used here is given in Appendix A of \citet{Couturier2022}. If the Earth is spherical, then its potential is radial and we take $\zeta(\theta,\varphi)=\Rearth$, that is, $\epsilon_{lm}=0$ for all $l$ and $m$.

Similarly as for the geoid, we write the potential raised by the redistribution of mass within the Earth as (\textit{e.g.} \citet{BoueCorreiaLaskar2019})
\begin{equation}
	\begin{split}\label{V_potential}
		&V(r,\theta,\varphi)=-\frac{\G\Mearth}{r}\left[1+\hat{v}(r,\theta,\varphi)\right]+V_\Omega(r,\theta),\\
		&\hat{v}(r,\theta,\varphi)=\sum_{l=2}^{+\infty}\sum_{m=-l}^l\left(\frac{\Rearth}{r}\right)^l\hat{V}_{lm}Y_{lm}(\theta,\varphi),
	\end{split}
\end{equation}
where $V_\Omega(r,\theta)=\Omega^2r^2\left(P_2(\cos\theta)-1\right)/3$ is the potential raised by the rotation itself. We denote $\Omega_\text{c}=$ \smash{$\left(\G\Mearth/\Rearth^3\right)^{1/2}$} the Keplerian frequency at Earth's surface. With this notation, the potential raised by the Earth deformed under its own rotation can be rewritten
\begin{equation}
	\begin{split}
		&V(r,\theta,\varphi)=-\frac{\G\Mearth}{r}\left[1+v(r,\theta,\varphi)\right]-\frac{1}{3}\Omega^2r^2,\\
		&v(r,\theta,\varphi)=\sum_{l=2}^{+\infty}\sum_{m=-l}^l\left(\frac{\Rearth}{r}\right)^lV_{lm}Y_{lm}(\theta,\varphi),
	\end{split}
\end{equation}
where $V_{lm}=\hat{V}_{lm}$ if $\left(l,m\right)\neq\left(2,0\right)$ and 
\begin{equation}\label{V20}
	V_{20}=\hat{V}_{20}-\frac{1}{3}\frac{\Omega^2}{\Omega_\text{c}^2}\frac{r^5}{\Rearth^5}.
\end{equation}
If we assume $h\ll1$ and $v\ll1$ (this is equivalent to $\Omega^2\ll\Omega_\text{c}^2$), then it is easy to verify from the definition of the geoid that (\citet{Wahr1996}, Sect. 4.3.1)
\begin{equation}\label{relation_geoid_potential}
	\epsilon_{lm}=V_{lm}\bigr|_{r=\Rearth}.
\end{equation}
This gives a relation between the figure of the Earth (the geoid) and the potential raised by the redistribution of mass. If we limit ourselves to the quadrupolar order and if we assume that the problem does not to depend on $\varphi$ (axisymmetry), then all the $V_{lm}$ and $\epsilon_{lm}$ vanish for $\left(l,m\right)\neq\left(2,0\right)$. For the fluid Earth, it can be shown that (\citet{Couturier2022}, Sect. 5.2.1; \citet{Wahr1996}, Eq. (4.24))
\begin{equation}\label{epsilon_20}
	\epsilon_{20}=-\frac{5}{6}\frac{\Omega^2}{\Omega_\text{c}^2}.
\end{equation}
The $J_2$ coefficient is defined as $J_2=-\hat{V}_{20}$ (with the convention of Appendix A of \citet{Couturier2022} for the spherical harmonics). For the fluid Earth, Eqs. (\ref{V20}), (\ref{relation_geoid_potential}) and (\ref{epsilon_20}) yield
\begin{equation}\label{J2}
	J_2=\frac{1}{2}\frac{\Omega^2}{\Omega_\text{c}^2}.
\end{equation}
According to Eq. (\ref{V_potential}), a moonlet orbiting the Earth at position $\vect{r}$ in the geocentric reference frame, feels, from the equatorial bulge, the potential per unit mass\footnote{Due to the axisymmetry, we can go to the geocentric reference frame by simply removing $V_\Omega$ in Eq. (\ref{V_potential}).}
\begin{equation}
	\begin{split}
	V_{J_2}=&\frac{\G\Mearth\Rearth^2}{r^3}J_2P_2(\cos\theta)\\
	=-&\frac{\G\Mearth\Rearth^2}{2r^5}J_2\left(r^2-3\left(\vect{k}\cdot\vect{r}\right)^2\right).
	\end{split}
\end{equation}
Writing $\vect{r}=x\vect{i}+y\vect{j}+z\vect{k}$, we have $\vect{k}\cdot\vect{r}=z$, and then using Eq. (\ref{acceleration}), the contribution of Earth's equatorial bulge to the acceleration of a moonlet takes the form
\begin{equation}
	\ddot{\vect{r}}=\frac{\G\Mearth\Rearth^2J_2}{r^5}\left[\frac{15z^2-3r^2}{2r^2}\vect{r}-3z\vect{k}\right].
\end{equation}
The user chooses the sideral rotation period of the Earth (or central body) in the parameter file of Ncorpi$\OO$N. Then, Eq. (\ref{J2}) and the fluid approximation are used to determine the $J_2$ of the central body.

\subsection{Interaction with the Sun}

The interaction between a moonlet, located at $\vect{r}$, and the Sun, located at $\vect{r}_\odot$ in the geocentric reference frame can be taken into account in the model by adding to the moonlet the potential per unit mass
\begin{equation}
	V_\odot=-\G\Msun\left(\frac{1}{\left|\vect{r}-\vect{r}_\odot\right|}-\frac{\vect{r}\cdot\vect{r}_\odot}{r_\odot^3}\right).
\end{equation}
To the quadrupolar order, this gives
\begin{equation}
	V_\odot=-\frac{\G\Msun}{2r_\odot^3}\left(3\frac{\left(\vect{r}\cdot\vect{r}_\odot\right)^2}{r_\odot^2}-r^2\right).
\end{equation}
Equation (\ref{acceleration}) yields, for the acceleration of the moonlet
\begin{equation}
	\ddot{\vect{r}}=-\frac{\G\Msun}{r_\odot^3}\left(\vect{r}-3\frac{\vect{r}\cdot\vect{r}_\odot}{r_\odot^2}\vect{r}_\odot\right).
\end{equation}
For simplification, we assume that the Earth orbits the Sun on a circular trajectory. Without restraining the generality, we can assume that the Earth's longitude of the ascending node is $0$ (this can be achieved by simply putting the unit vector $\vect{i}$ of the geocentric frame towards the ascending node). Similarly, for a circular orbit, we can arbitrarily choose $\omega_\odot=0$. We denote $a_\odot$ the semi-major axis of the Earth orbit and $\epsilon$ the obliquity of the Earth. With these notations, the vector $\vect{r}_\odot$ reads
\begin{equation}
	\begin{split}
	&\vect{r}_\odot=
	\begin{pmatrix}x_\odot\\y_\odot\\z_\odot\end{pmatrix}=
	\begin{pmatrix}
		1&0&0\\
		0&\cos\epsilon&-\sin\epsilon\\
		0&\sin\epsilon&\cos\epsilon
	\end{pmatrix}\begin{pmatrix}
		a_\odot\cos\lambda_\odot\\a_\odot\sin\lambda_\odot\\0
	\end{pmatrix}\!\!\!\\
	&=\begin{pmatrix}
		a_\odot\cos\lambda_\odot\\a_\odot\sin\lambda_\odot\cos\epsilon\\a_\odot\sin\lambda_\odot\sin\epsilon
	\end{pmatrix},
	\end{split}
\end{equation}
where $\lambda_\odot=n_\odot t=t\sqrt{\G\Msun/a_\odot^3}$.

\section{Multipole moment $\vect{M}^{(n)}(\vect{s}_B)$ of a cell from those of its children}\label{multipole_from_children}

We give here for $n\leq5$ the expression of the multipole moment \smash{$\vect{M}^{(n)}(\vect{s}_B)$} of a parent cell (Eq. \ref{multipole_moment}) from the multipole moments \smash{$\vect{m}^{(n)}(\vect{s}_b)$} of its children cells. The notations $\vect{s}_B$ and $\vect{s}_b$ are the expansion centers of the parent and of one of its children. We denote $\vect{\Lambda}=\vect{s}_B-\vect{s}_b=\left(\Lambda_1,\Lambda_2,\Lambda_3\right)$ and we assume that the expansion centers are the barycentres, leading to the simplification \smash{$\vect{M}^{(1)}(\vect{s}_B)=$} \smash{$\vect{m}^{(1)}(\vect{s}_b)=\vect{0}$}. The contribution \smash{$\vect{M}^{(n)}_{b\to B}$} from child $b$ to the multipole moment of its parent $B$ is given by
\begin{equation}
	\!\!\!\!\!\!\!\!\!\!\!\!\!\!\!\!\!\!\!\!\!\!\!\!\!\!\!\!\!\!\!\!\!\!\!\!\!\!\!\!\!\!\!\!\!\!\!\!\!\!\!\!\!\!\!\!\!\!\!\!\!\!\!\!\!\!\!\!\!\!\!\!\!\!\!\!\!\!\!\!\!\!\!\!\!\!\!\!\!\!\!\!\!\!\!\!\!\!\!\!\!\!\!\!\!\!\!\!\!\!M^{(0)}_{b\to B}=m^{(0)},
\end{equation}
\begin{equation}
	\!\!\!\!\!\!\!\!\!\!\!\!\!\!\!\!\!\!\!\!\!\!\!\!\!\!\!\!\!\!\!\!\!\!\!\!\!\!\!\!\!\!\!\!\!\!\!\!\!\!\!\!\!\!\!\!\!\!\!\!\!\!\!\!\!\!\!\!\!\!\!\!\vect{M}_{b\to B}^{(2)}=\vect{m}^{(2)}+m^{(0)}\vect{\Lambda}\otimes\vect{\Lambda},
\end{equation}
\begin{equation}
	\begin{split}
	&\!\!\!\!\!\!\!\!\!\!\!\!\!\!\!\!\!\!\left[\vect{M}_{b\to B}^{(3)}\right]_{ijk}=m_{ijk}^{(3)}\\
	&\!\!\!\!\!\!\!\!\!\!\!\!\!\!\!\!\!\!-\left(m_{ij}^{(2)}\Lambda_k+m_{ik}^{(2)}\Lambda_j+m_{jk}^{(2)}\Lambda_i\right)-m^{(0)}\Lambda_i\Lambda_j\Lambda_k,
	\end{split}
\end{equation}
\begin{equation}
	\begin{split}
		&\!\!\!\!\!\!\left[\vect{M}_{b\to B}^{(4)}\right]_{ijkl}=m_{ijkl}^{(4)}-\left(m_{ijk}^{(3)}\Lambda_l+m_{ijl}^{(3)}\Lambda_k+m_{ikl}^{(3)}\Lambda_j\right.\\
		&\!\!\!\!\!\! \left. +\;m_{jkl}^{(3)}\Lambda_i\right)+\left(m_{ij}^{(2)}\Lambda_k\Lambda_l+m_{ik}^{(2)}\Lambda_j\Lambda_l+m_{il}^{(2)}\Lambda_j\Lambda_k\right.\\
		&\!\!\!\!\!\! \left. +\;m_{jk}^{(2)}\Lambda_i\Lambda_l+m_{jl}^{(2)}\Lambda_i\Lambda_k+m_{kl}^{(2)}\Lambda_i\Lambda_j\right)+m^{(0)}\Lambda_i\Lambda_j\Lambda_k\Lambda_l,\!\!\!\!\!\!\!\!\!
	\end{split}
\end{equation}
\begin{equation}
	\begin{split}
		&\!\!\!\!\!\!\!\;\!\left[\vect{M}_{b\to B}^{(5)}\right]_{ijklm}=m_{ijklm}^{(5)}-\left(m_{ijkl}^{(4)}\Lambda_m+m_{ijkm}^{(4)}\Lambda_l\right.\\
		&\!\!\!\!\!\!\!\;\! \left. +\;m_{ijlm}^{(4)}\Lambda_k+m_{iklm}^{(4)}\Lambda_j+m_{jklm}^{(4)}\Lambda_i\right)+\left(m_{ijk}^{(3)}\Lambda_l\Lambda_m\right.\\
		&\!\!\!\!\!\!\!\;\! \left. +\;m_{ijl}^{(3)}\Lambda_k\Lambda_m+m_{ijm}^{(3)}\Lambda_k\Lambda_l+m_{ikl}^{(3)}\Lambda_j\Lambda_m+m_{ikm}^{(3)}\Lambda_j\Lambda_l\right.\!\!\!\\
		&\!\!\!\!\!\!\!\;\! \left. +\;m_{ilm}^{(3)}\Lambda_j\Lambda_k+ m_{jkl}^{(3)}\Lambda_i\Lambda_m+m_{jkm}^{(3)}\Lambda_i\Lambda_l+m_{jlm}^{(3)}\Lambda_i\Lambda_k\right.\!\!\!\\
		&\!\!\!\!\!\!\!\;\! \left. +\;m_{klm}^{(3)}\Lambda_i\Lambda_j\right)-\left(m_{ij}^{(2)}\Lambda_k\Lambda_l\Lambda_m+m_{ik}^{(2)}\Lambda_j\Lambda_l\Lambda_m\right.\\
		&\!\!\!\!\!\!\!\;\! \left. +\; m_{il}^{(2)}\Lambda_j\Lambda_k\Lambda_m+m_{im}^{(2)}\Lambda_j\Lambda_k\Lambda_l+m_{jk}^{(2)}\Lambda_i\Lambda_l\Lambda_m\right.\\
		&\!\!\!\!\!\!\!\;\! \left. +\;m_{jl}^{(2)}\Lambda_i\Lambda_k\Lambda_m+m_{jm}^{(2)}\Lambda_i\Lambda_k\Lambda_l+m_{kl}^{(2)}\Lambda_i\Lambda_j\Lambda_m\right.\\
		&\!\!\!\!\!\!\!\;\! \left. +\;m_{km}^{(2)}\Lambda_i\Lambda_j\Lambda_l+m_{lm}^{(2)}\Lambda_i\Lambda_j\Lambda_k\right)-m^{(0)}\Lambda_i\Lambda_j\Lambda_k\Lambda_l\Lambda_m,\!\!\!
	\end{split}
\end{equation}
where $\left(i,j,k,l,m\right)\in\left\lbrace1,2,3\right\rbrace^5$. The multipole moment of the parent is then obtained by summing over its children
\begin{equation}
	\vect{M}^{(n)}(\vect{s}_B)=\sum_{\text{children }b}\vect{M}^{(n)}_{b\to B}.
\end{equation}

\section{Conservation of the angular momentum upon fragmenting or merging impact}\label{angular_momentum_conservation}

We present here the method that Ncorpi$\OO$N uses to preserve the angular momentum up to machine precision, when the user requires so. We recall that
\begin{equation}
	\vect{v}_\text{cm}=\frac{m_1}{M}\vect{v_1}+\frac{m_2}{M}\vect{v_2}\quad\text{and}\quad\vect{r}_\text{cm}=\frac{m_1}{M}\vect{r_1}+\frac{m_2}{M}\vect{r_2},
\end{equation}
are the velocity and position of the center of mass of the colliding pair, whereas 
\begin{equation}
	\vect{G}=m_1\vect{r}_1\times\vect{v}_1+m_2\vect{r}_2\times\vect{v}_2,
\end{equation}
is the angular momentum to be conserved.

\subsection{Case of a merger}

If the collision results in a merger, then the outcome is a single moonlet of mass $M=m_1+m_2$.
The conservation of the total angular momentum reads
\begin{equation}\label{total_angular_momentum_merger}
	\vect{G}=M\tilde{\vect{r}}\times\tilde{\vect{v}}.
\end{equation}
Equation (\ref{total_angular_momentum_merger}) only has solutions if $\tilde{\vect{r}}$ is perpendicular to $\vect{G}$. Therefore, we write
\begin{equation}
	\tilde{\vect{r}}=\vect{r}_\text{cm}+\delta\tilde{\vect{r}},
\end{equation}
and we choose the smallest possible value of $\delta\tilde{\vect{r}}$ that verifies
\begin{equation}\label{delta_r}
	\vect{G}\cdot\tilde{\vect{r}}=\vect{G}\cdot\delta\tilde{\vect{r}}+\frac{m_1m_2}{M}\vect{r}_2\cdot\Delta\vect{r}\times\Delta\vect{v}=0.
\end{equation}
Equation (\ref{delta_r}) is of the form $\vect{\mathfrak{a}}\cdot\vect{w}=\mathfrak{b}$ with unknown $\vect{w}=\delta\tilde{\vect{r}}$. We are lead to minimize $\left|\vect{w}\right|^2$ under the constraint $\vect{\mathfrak{a}}\cdot\vect{w}=\mathfrak{b}$. We write
\begin{equation}
	\mathcal{L}(\lambda,\vect{w})=\left|\vect{w}\right|^2+\lambda\left(\vect{\mathfrak{a}}\cdot\vect{w}-\mathfrak{b}\right),
\end{equation}
where $\lambda$ is a Lagrange multiplier. The gradient of $\mathcal{L}$ vanishes when $\vect{w}=\mathfrak{b}\vect{\mathfrak{a}}/\mathfrak{a}^2$, and therefore we take
\begin{equation}
	\delta\tilde{\vect{r}}=\frac{m_1m_2}{M}\vect{r}_2\cdot\left(\Delta\vect{v}\times\Delta\vect{r}\right)\frac{\vect{G}}{G^2}.
\end{equation}
Once $\tilde{\vect{r}}$ is known, Eq. (\ref{total_angular_momentum_merger}) has the form $\vect{a}\times\vect{w}=\vect{b}$ with unknown $\vect{w}=\tilde{\vect{v}}$. Since $\vect{a}\cdot\vect{b}=0$, this equation has solutions given by\footnote{\label{sol_axv=b}This comes from $\vect{a}\times\left(\vect{b}\times\vect{a}\right)=a^2\vect{b}-\left(\vect{a}\cdot\vect{b}\right)\vect{a}$.} $\vect{w}=\left(\vect{b}\times\vect{a}\right)/a^2+\alpha\vect{a}$ for any $\alpha\in\mathbb{R}$. Therefore, we take 
\begin{equation}
	\tilde{\vect{v}}=\frac{1}{M\tilde{r}^2}\vect{G}\times\tilde{\vect{r}}+\alpha\tilde{\vect{r}},
\end{equation}
where we choose the real number $\alpha$ in order to minimize $\left|\tilde{\vect{v}}-\vect{v}_\text{cm}\right|$. We have 
\begin{equation}
	\left|\tilde{\vect{v}}-\vect{v}_\text{cm}\right|^2=\alpha^2\tilde{r}^2-2\alpha\tilde{\vect{r}}\cdot\vect{v}_\text{cm}+K,
\end{equation}
where $K$ does not depend on $\alpha$, and the minimal value of $\left|\tilde{\vect{v}}-\vect{v}_\text{cm}\right|$ is thus reached at $\alpha=\left(\tilde{\vect{r}}\cdot\vect{v}_\text{cm}\right)/\tilde{r}^2$. Finally, we achieve the conservation of the total angular momentum by giving to the unique moonlet resulting from the merger the position and velocity
\begin{equation}
	\left\lbrace\begin{matrix}
		\displaystyle{\tilde{\vect{r}}=\vect{r}_\text{cm}+\frac{m_1m_2}{M}\vect{r}_2\cdot\left(\Delta\vect{v}\times\Delta\vect{r}\right)\frac{\vect{G}}{G^2},}\\
		\displaystyle{\tilde{\vect{v}}=\frac{1}{M\tilde{r}^2}\vect{G}\times\tilde{\vect{r}}+\frac{\tilde{\vect{r}}\cdot\vect{v}_\text{cm}}{\tilde{r}^2}\tilde{\vect{r}}.\quad\quad\quad\;\;\;\,}
	\end{matrix}\right.
\end{equation}

\subsection{Case of a fragmentation}

If the collision results in a full fragmentation ($\tilde{m}_2\geq m^{(0)}$), then the conservation of the total angular momentum reads
\begin{equation}\label{total_angular_momentum_fragmentation}
	\vect{G}=\tilde{m}\tilde{\vect{r}}\times\tilde{\vect{v}}+\tilde{m}_2\sum_{k=1}^{\tilde{N}}\left(\tilde{\vect{r}}+\tilde{\vect{r}}_k'\right)\times\left(\tilde{\vect{v}}+\tilde{\vect{v}}_k'\right).
\end{equation} 
In the case of a partial fragmentation (\smash{$\tilde{m}_2< m^{(0)}\leq\check{m}$}), the tail is reunited into a single moonlet and the sum in Eq. (\ref{total_angular_momentum_fragmentation}) has only one term. We define\footnote{For a partial fragmentation, the sums are reduced to one term and $\tilde{m}_2$ has to be replaced by $\check{m}$.}
\begin{equation}\label{gsu}
	\begin{split}
	&\tilde{\vect{g}}=\tilde{m}_2\sum_{k=1}^{\tilde{N}}\tilde{\vect{r}}_k'\times\tilde{\vect{v}}_k',\\
	&\tilde{\vect{s}}=\tilde{m}_2\sum_{k=1}^{\tilde{N}}\tilde{\vect{r}}_k',\\
	&\tilde{\vect{u}}=\tilde{m}_2\sum_{k=1}^{\tilde{N}}\tilde{\vect{v}}_k',
	\end{split}
\end{equation}
and Eq. (\ref{total_angular_momentum_fragmentation}) can be rewritten
\begin{equation}
	\vect{G}=M\tilde{\vect{r}}\times\tilde{\vect{v}}+\tilde{\vect{r}}\times\tilde{\vect{u}}+\tilde{\vect{s}}\times\tilde{\vect{v}}+\tilde{\vect{g}},
\end{equation}
with unknowns $\tilde{\vect{r}}$ and $\tilde{\vect{v}}$. If $\tilde{\vect{r}}$ is known, then $\tilde{\vect{v}}$ is given by the equation
\begin{equation}\label{eq_for_v_tilde}
	\begin{split}
	&\vect{a}\times\tilde{\vect{v}}=\vect{b},\quad\text{where}\\
	&\vect{a}=M\tilde{\vect{r}}+\tilde{\vect{s}}\;\text{ and }\;\vect{b}=\vect{G}-\tilde{\vect{r}}\times\tilde{\vect{u}}-\tilde{\vect{g}}.
	\end{split}
\end{equation}
Equation (\ref{eq_for_v_tilde}) only has solutions if $\vect{a}\cdot\vect{b}=0$, and we first constrain $\tilde{\vect{r}}$ with the equation $\vect{a}\cdot\vect{b}=0$. Then, we obtain $\tilde{\vect{v}}$ from Eq. (\ref{eq_for_v_tilde}). There are infinitely many choices for both $\tilde{\vect{r}}$ and $\tilde{\vect{v}}$, and in each case we choose them in order to be as close as possible from the conservation of the total momentum, that is, as close as possible to
\begin{equation}\label{Conservation of total momentum}
	\begin{split}
		&\tilde{m}\tilde{\vect{r}}+\tilde{m}_2\sum_{k=1}^{\tilde{N}}\left(\tilde{\vect{r}}+\tilde{\vect{r}}_k'\right)=M\tilde{\vect{r}}+\tilde{\vect{s}}=M\vect{r}_\text{cm},\\
		&\tilde{m}\tilde{\vect{v}}+\tilde{m}_2\sum_{k=1}^{\tilde{N}}\left(\tilde{\vect{v}}+\tilde{\vect{v}}_k'\right)=M\tilde{\vect{v}}+\tilde{\vect{u}}=M\vect{v}_\text{cm}.
	\end{split}
\end{equation}
In order to determine $\tilde{\vect{r}}$, we thus write $M\tilde{\vect{r}}+\tilde{\vect{s}}=M\left(\vect{r}_\text{cm}+\delta\tilde{\vect{r}}\right)$ and we choose the smallest $\delta\tilde{\vect{r}}$ that verifies $\vect{a}\cdot\vect{b}=0$. We have
\begin{equation}
	\begin{split}
	\vect{a}\cdot\vect{b}&=\left(M\tilde{\vect{r}}+\tilde{\vect{s}}\right)\cdot\left(\vect{G}-\tilde{\vect{g}}\right)+\tilde{\vect{r}}\cdot\left(\tilde{\vect{s}}\times\tilde{\vect{u}}\right)\\
	&=\left(\vect{r}_\text{cm}+\delta\tilde{\vect{r}}\right)\cdot\left(M\vect{G}-M\tilde{\vect{g}}+\tilde{\vect{s}}\times\tilde{\vect{u}}\right)=0.
	\end{split}
\end{equation}
We are left to minimize $\left|\delta\tilde{\vect{r}}\right|$ under a constraint of the form $\vect{\mathfrak{a}}\cdot\delta\tilde{\vect{r}}=\mathfrak{b}$. This was already done in the merger case with the theory of Lagrange multiplier and we have
\begin{equation}
	\begin{split}
	&\delta\tilde{\vect{r}}=\frac{\mathfrak{b}\vect{\mathfrak{a}}}{\mathfrak{a}^2}=-\frac{\left(\vect{r}_\text{cm}\cdot\vect{\mathfrak{a}}\right)\vect{\mathfrak{a}}}{\mathfrak{a}^2},\quad\text{where}\\
	&\vect{\mathfrak{a}}=M\left(\vect{G}-\tilde{\vect{g}}\right)+\tilde{\vect{s}}\times\tilde{\vect{u}}.
	\end{split}
\end{equation}
Now that $\tilde{\vect{r}}$ is known, we can obtain $\tilde{\vect{v}}$ from Eq. (\ref{eq_for_v_tilde}). The solutions of Eq. (\ref{eq_for_v_tilde}) are given by\textsuperscript{\ref{sol_axv=b}}
\begin{equation}
	\tilde{\vect{v}}=\frac{\vect{b}\times\vect{a}}{a^2}+\alpha\vect{a},
\end{equation}
where $\alpha\in\mathbb{R}$. We choose for the real number $\alpha$ the value that is closest from preserving the total momentum, that is, we choose the value of $\alpha$ that minimizes \smash{$\left|M\left(\tilde{\vect{v}}-\vect{v}_\text{cm}\right)+\tilde{\vect{u}}\right|$} (see Eq. (\ref{Conservation of total momentum})). We have 
\begin{equation}
	\begin{split}
	&\frac{1}{M^2}\left|M\left(\tilde{\vect{v}}-\vect{v}_\text{cm}\right)+\tilde{\vect{u}}\right|^2\\
	=\;&a^2\alpha^2-2\alpha\left(\vect{v}_\text{cm}-\frac{\tilde{\vect{u}}}{M}\right)\cdot\vect{a}+K,
	\end{split}
\end{equation}
where $K$ does not depend on $\alpha$ and therefore, we choose
\begin{equation}
	\alpha=\frac{\left(\vect{v}_\text{cm}-\tilde{\vect{u}}/M\right)\cdot\vect{a}}{a^2}.
\end{equation}
We uniquely determined $\tilde{\vect{r}}$ and $\tilde{\vect{v}}$ in such a way that the total angular momentum is conserved upon impact up to machine precision, whether the collision results in a merger or in a fragmentation.


\bibliographystyle{elsarticle-num-names} 
\bibliography{ncorpion.bib}





\end{document}